# A simple model for the outcomes of collisions between exhaled aerosol droplets and airborne particulate matter: Towards an understanding of the influence of air pollution on airborne viral transmission

**Andrew N. Round**
**School of Pharmacy, University of East Anglia, Norwich Research Park, Norwich, UK, NR4 7TJ**
a.round@uea.ac.uk

## Abstract
A model that predicts the outcome of collisions between droplets and particles in terms of the distribution of the droplet volume post-collision is lacking, in contrast to the case for droplet-droplet interactions. Taking existing models that successfully predict the outcomes (coalescence, stretching or reflexive separation) and post-separation characteristics (sizes, numbers and velocities of the resulting droplets) of droplet-droplet collisions and adapting them to take into account an inextensible, non-deformable particle with varying wettability characteristics, a new model is presented for droplet-particle collisions. The predictions of the new model agree well with experimental observations of droplet-particle collisions in low-viscosity regimes.

The model is then applied to the case of collisions between respiratory aerosols generated by breath, speech, cough and sneeze and ambient airborne particulate material (PM) in order to assess the potential contribution of these interactions to the enhanced transmission of pathogens contained in the aerosol, including COVID-19. The results show that under realistic conditions it is possible for aerosol-PM collisions to enrich the pathogen content of smaller (and so more persistent) aerosol fractions, and to transfer pathogens to the surface of PM particles that can travel deep into the respiratory tract. In the context of better knowledge of the size and velocity distributions of respiratory aerosols, this model may be used to predict the extent to which high ambient PM levels may contribute to airborne infection by pathogens such as COVID-19.

## Introduction
Interactions between droplets and particles have been studied for many years, across different disciplines. Collisions between particles and/or droplets in the atmosphere have generally been considered in the context of the agglomeration of materials to form particulate matter, where low-velocity collisions lead to coalescence [1]. Interactions between fluid droplets and solid curved surfaces are ubiquitous in nature and industry, from ice nucleation on power lines [2] to the beading of rain droplets on plant leaves [3], so that understanding the physics of their interactions is of great importance.

Much interest in these interactions arises from the need to optimise industrial processes such as catalytic cracking, where the focus is on optimising rates of heat transfer amid concerns over the vaporisation of the fluid droplets leading to non-wetting of the catalytic particles. Other processes dependent on heat transfer during droplet-particle interactions include many food and pharmaceutical production methods such as spray drying [4], freeze drying [5] and melt extrusion [6]. Thus much of the literature on droplet-particle interactions focuses on the outcomes of these interactions from the perspective of changes in heat. At high energies and/or temperatures, these interactions often result in the vaporisation or shattering of droplets during collision and take place in energetic regimes displaced from those that are relevant to aerosol-pollutant interactions. A smaller literature is concerned more specifically with collisions between droplets of fluids (mainly water) and spherical particles in the micron to millimetre size range and at speeds in the 1-10 $m.s^{-1}$ range. Drawing on a wider literature on droplet-droplet interactions [7-12], which show how collisions can lead to outcomes such as bouncing, coalescence and stretching and reflexive separation, with the generation of satellite droplets, experimental studies [13-19] have sought to characterise the outcomes of similar droplet-particle interactions in similar terms, but a predictive model to characterise the fate of fluid in a droplet that collides with a particle is missing.

A new field in which droplet-particle interactions may have significant consequences on public health is in the transmission of airborne pathogens in areas with high particulate pollution. Since the beginning of the COVID-19 pandemic, evidence of a link between high levels of airborne particulate material (PM) and increased incidence and severity of COVID-19 has accrued [20-22]. These findings are not unexpected, given that these studies complement previous work linking PM to other airborne viruses such as SARS and influenza [23-25]. As a recent review [22] summarises, the causative factors in this relationship may include comorbidities arising from both chronic and acute health effects of high levels of airborne pollution, meteorological factors such as temperature, humidity and UV levels [26], or there may be physical and chemical interactions between airborne pollutants and exhaled aerosol droplets that increase the transmission of COVID-19. The same work finds that little attention has been paid to the last of these: mechanisms by which direct interactions between airborne pollutant particles and exhaled respiratory aerosol droplets containing virions might enhance the transmission of airborne viruses.

In this work, the interactions between virion-containing respired droplets and ambient pollutant particles that may result in an enhancement in the transmission of the virus through aerosol inhalation are considered. The leading hypothesis behind this effort is that collisions between droplets and particles may result in the splitting of droplets, together with the possible transfer of virions to the particle surface. Since PM particles are generally smaller than aerosol droplets, and since the 'daughter' droplets resulting from the splitting of droplets during collisions are necessarily smaller, the proposition is that the virions coughed, sneezed or breathed out by an infected person may, following interaction with airborne pollutant particles, be enriched in smaller droplets and particles, and thus: persist in the air for longer, increasing the chances of transmissibility; be drawn deeper into the lung on inhalation and so be more likely to lead to infection; or remain viable for longer on the surface of PM particles than they would in the aerosol.

Taking experimentally observed values for the distributions of sizes of aerosol droplets and PM particles and the velocities of droplets generated during respiration processes, existing models for the outcomes of collisions between droplets are used to develop a new predictive model for the outcomes of collisions between droplets and particles of these sizes and relative velocities, to assess the degree of enhancement of transmissibility that results from these interactions. The model may find application in related areas of study where droplet-particle interactions are important.

### 2.1 Droplet-droplet collision dynamics

Several models and experimental investigations of the outcomes of collisions between droplets have been studied [7-12]. A key finding has been that there are conditions under which separation of the two interacting droplets occurs after collision, and others where coalescence occurs. A threshold value, $E_{coal}$, can be determined to predict the outcomes following a collision. During a separation, liquid from the larger droplet may be transferred to the smaller, and smaller satellite droplets may be created. The probability of separation is determined by the balance of kinetic and surface energies acting on the droplets and so depends ultimately on the relative sizes and velocities of the droplets upon collision.

Pioneering work by O'Rourke [7] described a predictive model for collisions between droplets of diameters $d_1$ and $d_2$ on the basis of three dimensionless parameters, as defined in equations (1-3): the Weber number We, the droplet size ratio $\Delta$ and an impact parameter named x or b in different studies. Here we will set out the basic model, discuss its application to particle-droplet interactions and explore the consequences in terms of the number and sizes of post-collision droplets.

$$\Delta = \frac{d_1}{d_2} \tag{1}$$

$$b = \frac{2B}{(d_1 + d_2)} = \sin\theta \tag{2}$$

$$We = \frac{\rho d_1 |\vec{u}_{rel}|^2}{\sigma} \tag{3}$$

Two droplets (smaller droplet with diameter $d_1$ and larger droplet with diameter $d_2$, with droplet size ratio $\Delta$ defined in equation (1)) can collide at relative velocity $\vec{u}_{rel}$ in a range of configurations which may be characterized according to the scheme in figure 1. The impact parameter b is defined in eq. (2), where B is defined as the distance between the centres of the colliding droplets normal to the direction of the relative velocity vector $\vec{u}_{rel}$, as depicted in Figure 1. It can be seen that the limits of the value of B are 0 and $|(d_1 + d_2)/2|$, so that the impact parameter b can take values between 0 (representing a 'head-on' collision) and 1 (representing the finest glancing contact).

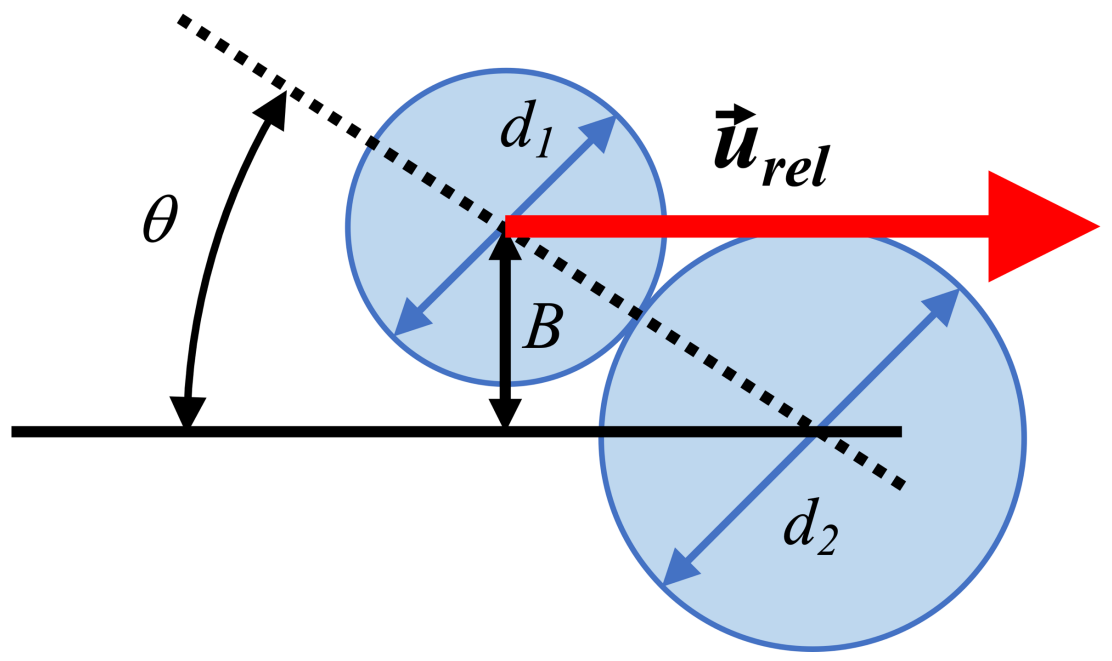

**Figure 1:** Scheme depicting the parameters used to determine the impact parameter *b*.

Finally, the Weber number We is defined by equation (3), where $\rho$ is the density and $\sigma$ is the surface tension of the droplet medium. The Weber number represents the relative importance of the droplet's inertia over its surface tension, or alternatively is proportional to the ratio of the kinetic and surface energies of the droplet, such that a high value of We favours the separation of the droplet(s) on contact, while a low value favours coalescence. O'Rourke [7] derived an expression for the value of b at the boundary between separation and coalescence, $E_{coal}$ (4):

$$E_{coal} = \min\left[1, \frac{2.4 f(\gamma)}{We}\right] \tag{4}$$

Where $f(\gamma) = \gamma^3 - 2.4\gamma^2 + 2.7\gamma$ and $\gamma = 1/\Delta$. Thus, for a binary droplet collision defined by values for $\Delta$, b and We, a regime plot for the separation and coalescence outcomes can be produced: an example is shown in Figure 2, where $\Delta = 0.5$, We varies from 0 to 100 and b varies from 0 to 1. This plot, and the discussion above, neglects the exceptional case where for very glancing collisions (high b) at the lowest values of We, the droplets bounce off each other without either coalescing or separating. This case will not be treated further in this work.

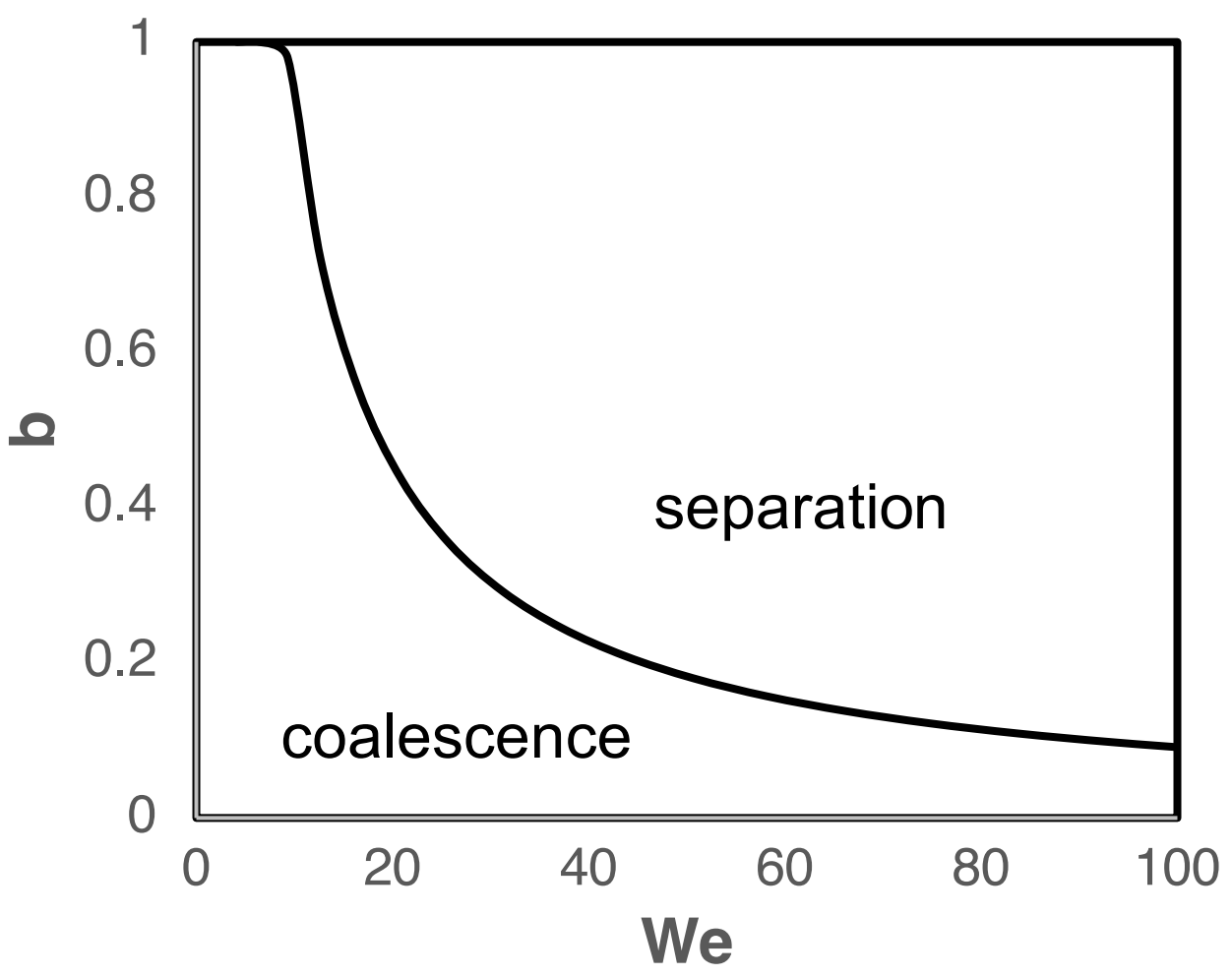

**Figure 2**: We-b regime plot showing the boundary between separation and coalescence for a system of colliding water droplets with $\Delta = 0.5$.

Ashgriz and Poo [8] extended O'Rourke's work following experimental observations that suggested that O'Rourke's model did not adequately predict separation and failed to account for the different types of post-collision droplet separation that they and others observed. They produced a theoretical model which accounted for their experimental data by explicitly considering the changes in kinetic and surface energies of droplets during the post-collision process. In particular, they identified two classes of separation: 'reflexive' separation typically following collisions with low b values, and 'stretching' separation following high-b collisions. The terms describing the outcome of these collisions and their post-separation characteristics are presented below and their derivation is recapitulated in the **Supporting Information**.

For the reflexive separation, at low values of the impact parameter (close to 'head-on') the coalescence boundary $We_{Rdd}$ is given as:

$$We_{Rdd} = 3\Delta(1+\Delta^3)^2\left[\frac{7(1+\Delta^3)^{2/3} - 4(1+\Delta^2)}{(\Delta^6\eta_1 + \eta_2)}\right] \tag{5}$$

And for the stretching separation at high values of b ('glancing blows') the boundary $We_{Sdd}$ is given as:

$$We_{Sdd} = \frac{4(1+\Delta^3)^2[3(1+\Delta)(1-b)(\Delta^3\phi_1 + \phi_2)]^{1/2}}{\Delta^2[(1+\Delta^3) - (1-b^2)(\phi_1 + \Delta^3\phi_2)]} \tag{6}$$

With the post-separation characteristics ($d_{2a}$, the diameter of the droplet after collision, $\vec{\boldsymbol{u}}_{\boldsymbol{na}}$, the post-collision velocities of the particle and droplet and $N_{sa}$, $d_{sa}$ and $\vec{\boldsymbol{u}}_{\boldsymbol{sa}}$, the number, diameter (assumed to be equal for a given separation) and velocity of satellite droplets generated by the separation) determined by Ko and Ryou as:

| **Stretching separation** | **Reflexive separation** | |
|---|---|---|
| $d_{2a} = (1 - C_{VS}\phi_2)^{1/3} d_2$ | $d_{2a} = (1 - C_{VR})^{1/3} d_2$ | (7a, b) |

$$\vec{u}_{na} = \vec{u}_n \qquad \boldsymbol{\vec{u}_{na}} = \frac{\sqrt{\mathbf{2}}}{\mathbf{2}}\boldsymbol{\vec{u}_n} \tag{8a, b}$$

$$N_{sa} = \left[\frac{SE_s}{\sigma\pi C_{VS}^{2/3}(\Delta^3\phi_1 + \phi_2)^{2/3} d_2^2}\right]^3 \qquad N_{sa} = \left[\frac{SE_s}{\sigma\pi C_{VR}^{2/3}(d_1^3 + d_2^3)^{2/3}}\right]^3 \tag{9a ,b}$$

$$d_{sa} = \left[\frac{C_{VS}(\Delta^3\phi_1 + \phi_2)}{N_{sa}}\right]^{1/3} d_2 \qquad d_{sa} = \left[\frac{C_{VR}(d_1^3 + d_2^3)}{N_{sa}}\right]^{1/3} \tag{10a, b}$$

$$\vec{u}_{sa} = \frac{\Delta^3(\phi_1 - \phi_2)}{(1+\Delta^3)(\Delta^3\phi_1 + \phi_2)}\vec{u}_{rel} \qquad \vec{u}_{sa} = \frac{d_1^3\vec{u}_1 + d_2^3\vec{u}_2}{d_1^3 + d_2^3} \tag{11a ,b}$$

All terms in eqs (5-11) are defined in the **Supporting Information**.

Figure 3(a, b) illustrates the collision and breakup processes for reflexive and stretching separations, following the collision and break-up captured in high-speed photographs shown by Ashgriz and Poo [8]. Figure 3(c) shows the results of equations (5) and (6), showing the regions of the regime plot where reflexive and stretching separation occur, for the same conditions used in Figure 2 for the O'Rourke separation boundary $E_{coal}$.

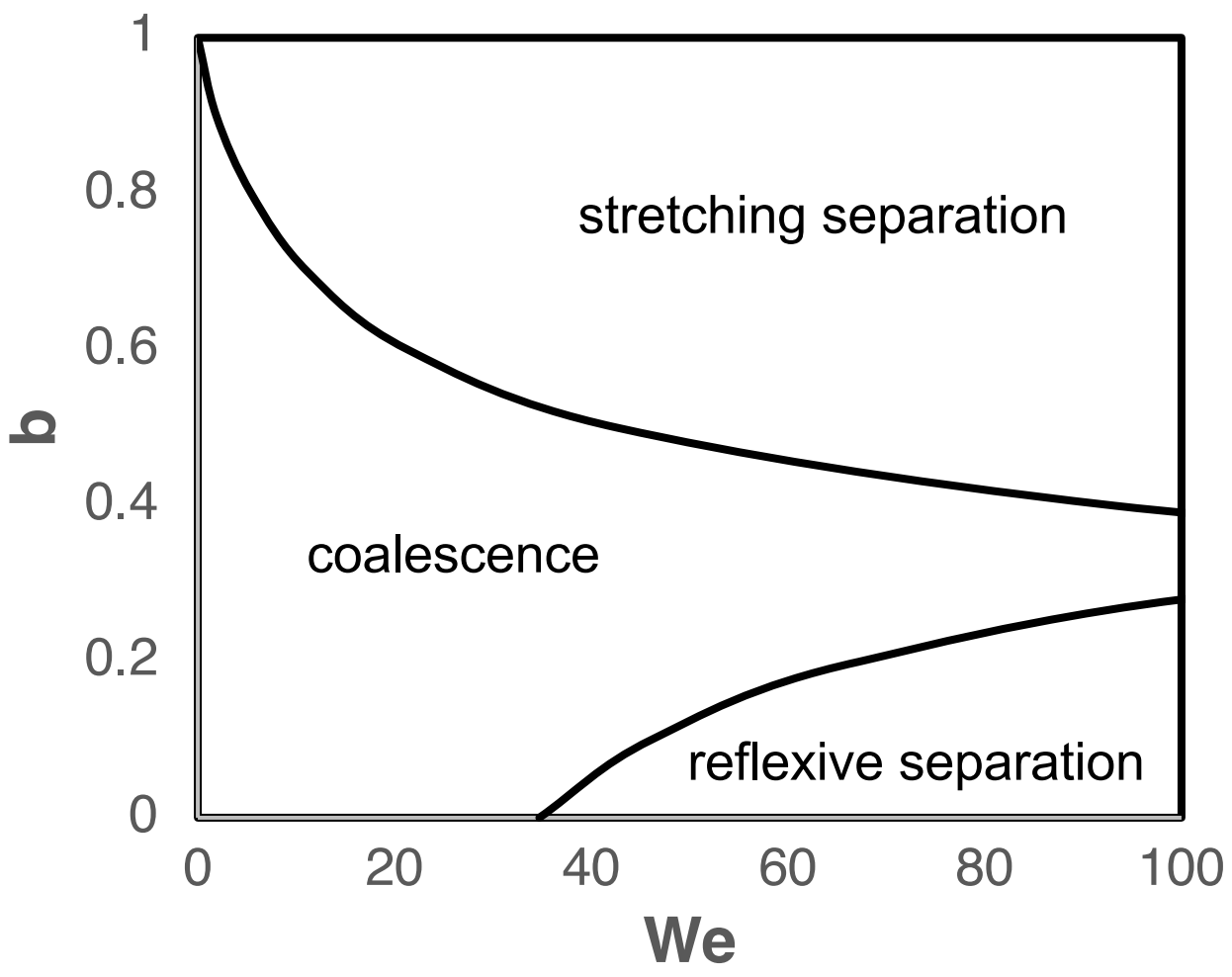

**Figure 3:** (a) sketch of the collision and breakup process for reflexive separation. (b) sketch of the collision and breakup process for stretching separation. (c) We-b regime plot showing stretching and reflexive separation for a system of water droplets with $\Delta = 0.5$.

It can be seen from comparison of figures 2 and 3(c) that the two models make different predictions about the conditions under which separation can occur following collisions between droplets, and experimental data collected by Ashgriz and Poo was better fit by their model [8]. It may be noted that no dissipative term is included in the energy considerations in the above treatment, which Ashgriz and Poo justified by consideration of the Laplace number for the systems under study. Indeed, Kuschel and Sommerfeld [10] found that, when comparing Ashgriz and Poo's model with that of Jiang et al [11] who predicted an energy dissipation coefficient $\alpha$ of 0.5 for collisions between water droplets, the former model gave the better fit. This result was reversed for liquids with higher viscosity, as will be touched on below.

## 2.2 Droplet-particle collision dynamics

The above analysis has only concerned collisions between liquid droplets with identical material properties. The question of characterizing collisions between liquid droplets and solid particles, while of great interest in a number of fields [19] besides its relevance to the case under consideration here, has received less attention. The first work to comprehensively study this question was that of Dubrovsky et al [13], who captured high speed film of collisions between small droplets (0.65-1.05 mm diameter, water:glycerol solutions with 0-91% glycerine, so that viscosity spanned two orders of magnitude) and large particles (3.2-8 mm diameter) at collision speeds between 7-10 $ms^{-1}$ (called hereafter p>d), and large droplets (2.9-5.6 mm diameter) and small particles (1 mm diameter, densities 2700-15200$kgm^{-3}$), at collision speeds between 3.4-12.8 $ms^{-1}$ (d>p). The data therefore covered values of $\Delta$ from 0.08 to 0.33 and We in the range $10^2$ - $10^4$ with collisions occurring at impact angles $\varphi$ between 0º and 60º (in the notation used previously, sin $\varphi$ is equivalent to b, the impact parameter). From their experimental data for the p>d case they determined the coalescence parameter $X$, being the point at which coalescence or breakup were equally likely (and so corresponding conceptually to O'Rourke's $E_{coal}$ for d-d collisions) and observed that $X$ increased (coalescence became more likely) with the size of the particle and with the viscosity of the droplet medium, while $X$ decreased as the contact angle increased (i.e. as b increased). This pattern of results follows the behaviour expected of a stretching separation in the terms described in the previous section. In the d>p experiments four different outcomes were observed: 'capture' of the particle by the droplet (coalescence); 'shooting through' where the droplet leaves the particle with a small number of droplet fragments; a subset of 'shooting through' where the particle leaves a hole in the droplet so that a hollow ring is formed; and finally 'destruction' where the droplet shatters into many fragments. The first outcome, coalescence, was favoured by head-on collision, low relative velocity, small values of $\Delta$, low particle density and high droplet viscosity. A lower density particle caused greater breakup of the droplet (so favouring 'destruction' over 'shooting through'), which goes against expectation but was attributed to a shorter contact time for the less dense particle, allowing less deformation of the droplet, while increasing relative collision velocity increased destruction. Capture was favoured by high liquid viscosity and greater size disparity between droplet and particle and disfavoured by denser particles and increased collision velocities.

Gac and Gradon [14] used a Lattice-Boltzmann method to simulate collisions between water droplets and solid spherical particles using the Weber number We, a droplet:particle diameter ratio $\Delta$, the contact angle $\theta$ and the Capillary number $Ca = \mu\vec{u}_{rel}d_d/\sigma d_p$ as controlling parameters. They described three classes of collision outcome: deposition (coalescence) at We = 0.6; 'ripping and coating', where the droplet engulfs the particle and departs at the other side, leaving a residual volume on the particle and generating a small number of satellite droplets, at We = 20; and 'skirt scattering' where the droplet forms a lamellar crown or skirt around the particle, which then breaks up into satellite droplets, at We = 70. These three outcomes appear qualitatively to resemble those described by Dubrovsky et al, with 'capture' equalling 'deposition', 'shooting through' resembling 'ripping and coating' and 'skirt scattering' resembling either the 'hollow ring' subset of 'shooting through' events or the 'shattering' outcome. Besides the dependence on We mentioned above, Gac and Gradon noted that at capillary numbers above 1 the droplet underwent deformation before contacting the particle, often leading to non-'head-on' collisions, while low contact angles (corresponding to hydrophilic particle surfaces) increased the probability of coalescence. Changing $\Delta$ influenced outcomes such as the amount of residual liquid left on the particle and the number and sizes of the satellite droplets but did not significantly change which of the collision outcomes occurred.

An experimental study of water droplet-glass particle interactions was carried out by Pawar et al [15], who mapped the results of collisions between 2.9 mm water droplets with spherical glass droplets of diameter 2.5 or 4 mm on to a regime plot of We vs b in the range We = 0-50 and b = 0-1. They observed coalescence and stretching separation with the production of satellite droplets, but not reflexive separation. Working within the paradigm of droplet-droplet collision models, they derived an empirical modification of O'Rourke's boundary parameter $E_{coal}$ that better fit their data and attributed the lack of observation of reflexive separation to the relatively narrow range of conditions studied, in particular with respect to contact angle.

Charalampous and Hardalupas [16] also studied the head-on impact of droplets on a larger particle affixed on a narrow rod and examined by high-speed photography the outcomes of the collisions at a range of velocities and droplet sizes. The arrangement of their experiment mitigated against determining how liquid leaves the particle following the evolution of the collision, and these authors were primarily interested in resolving details of the fluid motion during the initial collision, and the formation of a 'crown' on the particle surface. Within the range limited by the parametric values $92 < We < 1015$, $.0070 < Oh < 0.0089$ and $0.09 < \Delta < 0.55$, again they observed three classes of collision: deposition (coalescence), 'overpass' which resembles 'shooting through' (although the extent of liquid leaving the droplet post-collision is not quantified) and 'splashing' which appears similar to Gac and Gradon's 'skirt scattering'. Deposition was observed at We values $\leq 200$ and splashing at We values $> 400$, while behaviour in the intermediate state depended on the size of the droplet: for small droplets overpass collisions were observed while for larger droplets deposition continued to occur until We reached 400 and splashing occurred. Banitabaei and Amirfazli [17] extended this work for stationary particle and mobile droplets where d>p across a very wide range of values of We while recently Islamova et al [18] have carried out similar experiments to Pawar et al using metal particles and recording similar observations in We-b regime plots.

With regard to class of collision outcome and Weber number, Dubrovsky [13], Gac [14], Charalampous [16]and Banitabaei [17] all observed separation events following 'head-on' collisions, which would have been classed as 'reflexive separations' in Ashgriz and Poo's terms. Aside from the simulations of Gac and Gradon, separation events from head-on collisions occurred at rather high values of We ($We \geq 100$), which were beyond the regimes that Pawar [15] and Islamova [18] investigated. Indeed, the shattering/splashing event is phenomenologically distinct from the shooting-through/overpass event, with the latter appearing to more closely resemble reflexive separation. Dubrovsky, Pawar and Islamova all observed 'stretching separations' during off-centre collisions at generally lower We values ( $We \geq 10$).

### 2.3 The flow regimes for collisions between aerosol droplets and pollutant particles

Given the aim of this work is to deduce the distribution of liquid from an aerosol droplet following a collision with an airborne particle, the approach taken by Pawar et al [15], to modify the more developed framework for predicting droplet-droplet interactions to reflect the droplet-particle case, seems the most appropriate route to follow. Therefore, in order to more accurately predict the result of collisions between exhaled aerosol droplets and ambient PM, in the following section a simple modification of Ashgriz and Poo's [8], and Ko and Ryou's [9], models will be presented to account for the effect of replacing one liquid droplet in the collision with a solid particle. This new model will then be tested against the results of these previous experimental studies.

A first step in achieving this is to establish whether the flow regimes considered in the droplet-droplet collision models are appropriate for the scenario considered here. Experimental data and the values of parameters that describe the flow regimes in the systems studied are given for references [8], [13], [15] and [16] in Table 1, alongside the ranges of these values expected to apply in the aerosol droplet: PM system. The experimental data presented by Ashgriz and Poo [8] (Table 1) encompassed the ranges $5 < We < 109$, $200 < Re < 2000$, $5\times10^{-3} < Ca < 1\times10^{-4}$, and $7300 < Lp < 3.6\times10^{4}$. The range of Reynolds numbers (Re) explored in this data put the system largely into the Newtonian flow region where the drag coefficient is independent of the Reynolds number and inertia dominates, while the capillary number (Ca) being much smaller than 1 means that the deformation of the liquid in the droplet prior to collision is negligible. The values taken by the Laplace number (Lp) indicate that viscous dissipation would be expected to be small, so the model neglects this contribution. Later work [10, 11] found that dissipation becomes important for liquids with higher viscosity, and that the model that takes dissipative energy loss into account becomes the better predictor of the coalescence-separation boundary when $\mathrm{Ca} > 0.577$ and $\mathrm{Oh} > 0.115$ (i.e. $\mathrm{Lp} < 75$). A final consideration is the deposition efficiency, related to the Stokes number Stk, which characterises the extent to which droplets in low Reynolds regimes collide with obstructing particles or flow around them in eddies. In all cases given in Table 1, the deposition efficiency (the fraction of droplets that hit a target particle) is above 0.999.

The exhaled respiratory aerosol is a complex heterogeneous mixture of droplets with varying sizes and compositions, both of which evolve rapidly and significantly with time once exhaled [27-30]. Due to the wide range of different measurement methods employed, size distributions of the aerosol aften appear to be bimodal, though a recent review [27] suggests that this may be due to loss of resolution at particular intermediate size ranges. Overall the studies present a picture of aerosol size distribution which ranges from $< 1\ \mu$m to $> 1$ mm, with an apparent monotonic decrease in number (with an approximate $1/\sqrt{x}$ dependence) over more than 4 orders of magnitude from the smaller to the larger. The main component of the aerosol is water, with dissolved salts, pulmonary surfactants, proteins, mucus and potentially pathogens being present in varying amounts, together constituting the mucosalivary fluid [27]. Surface tension of dilute mucosalivary fluid has been found to be lower than that of pure water, at 0.04 [31], reflecting the surface-active character of many of the components, and it has been shown to exhibit shear-thinning properties, with viscosity varying from 2.5 to 14.2 mPa.s. These characteristics give rise to a very wide range of values of Re (1-23000), Ca ($4.5\times10^{-5}$ – 1000) and Oh (0.013 – 2.2) which overlap with the regimes described above, for which the droplet-droplet model is valid, but also include regimes beyond the validity of the Ashgriz-Poo model. The consequences of this will be considered below.

Ambient pollutant particulate matter is also a highly heterogeneous mixture with respect to size and composition [1] due to their diverse origins. Ambient atmospheric pollutant particulate matter (PM) content is monitored as a key indicator of air quality for environmental health and is typically classified according to size as $PM_{2.5}$ (particulate material with a maximum diameter of 2.5 $\mu$m) and $PM_{10}$ (for a maximum diameter of 10 $\mu$m). National governments and supranational organisations such as the EU and WHO mandate limits to concentrations of $PM_{2.5}$, $PM_{10}$ and key constituents of the PM (sulfates, nitrates and others) above which effects on public health are increased. Overall PM are therefore smaller than aerosol droplets, so that most interactions between aerosol droplets and PM will be those where the droplet is larger than the particle, or $d > p$. The distribution of sizes has similar characteristics to that of the aerosol droplets, which an approximate $1/\sqrt{x^3}$ dependence fairly represents. A study of urban and roadside PM composition in the UK [32] found the major components to be elemental and organic carbon, sulfate, nitrate, chloride, iron and calcium: in more rural environments PM content can be equally high with ammonium nitrates and sulfates derived from agriculture.

| **Table 1a. parameters** | **Symbol/ equation** | **units** | [8] | | [13], d > p | | [13], p > d | | [15] | | [16] | | This work | |
|---|---|---|---|---|---|---|---|---|---|---|---|---|---|---|
| velocity | $U_{rel}$ | $m.s^{-1}$ | 2 | 4 | 3.4 | 12.8 | 7 | 10 | 0.065 | 1.15 | 6 | 11 | 1 | 23 |
| droplet diameter | $d_d$ | m | $1.10^{-4}$ | $5.10^{-4}$ | $6.5.10^{-4}$ | $1.1.10^{-3}$ | $2.9.10^{-3}$ | $5.6.10^{-3}$ | $2.9.10^{-3}$ | | $1.7.10^{-4}$ | $2.8.10^{-4}$ | $1.10^{-6}$ | $1.10^{-3}$ |
| particle diameter | $d_p$ | m | -- | | $3.2.10^{-3}$ | $8.10^{-3}$ | $1.10^{-3}$ | | $2.5.10^{-3}$ | $4.10^{-3}$ | $5.10^{-4}$ | $5.10^{-3}$ | $1.10^{-8}$ | $1.10^{-5}$ |
| droplet density | $\rho_d$ | $kg.m^{-3}$ | 998 | | 998 | | 998 | | 998 | | 998 | | 998 | |
| particle density | $\rho_p$ | $kg.m^{-3}$ | -- | | 2700 | 15200 | 2700 | 15200 | 2526 | | 2526 | | 1100 | 13500 |
| air density | $\rho_g$ | $kg.m^{-3}$ | 1.225 | | 1.225 | | 1.225 | | 1.225 | | 1.225 | | 1.225 | |
| surface tension | $\sigma$ | $N.m^{-1}$ | 0.073 | | 0.073 | | 0.073 | | 0.073 | | 0.073 | | 0.04 | |
| contact angle | $\theta$ | º | -- | | 55 | 107 | 55 | 107 | 70 | 80 | 59 | 67 | 30 | 120 |
| droplet viscosity | $\mu$ | $kg.m^{-1}.s^{-1}$ | $1.10^{-3}$ | | $1.10^{-3}$ | 0.014 | $1.10^{-3}$ | 0.014 | $1.10^{-3}$ | | $1.10^{-3}$ | | $2.5.10^{-3}$ | $1.4.10^{-2}$ |
| air viscosity | $\mu_g$ | $kg.m^{-1}.s^{-1}$ | $1.8.10^{-5}$ | | $1.8.10^{-5}$ | | $1.8.10^{-5}$ | | $1.8.10^{-5}$ | | $1.8.10^{-5}$ | | $1.8.10^{-5}$ | |

**Table 1b. Dimensionless parameters**

| | | | | | | | | | | | | | |
|---|---|---|---|---|---|---|---|---|---|---|---|---|---|
| Weber number | $We = \frac{\rho_d d_d U_{rel}^2}{\sigma}$ | 5 | 109 | 795 | 2620 | 458 | 12543 | 0.2 | 53.1 | 84 | 463 | 0 | 13199 |
| Reynolds number | $Re = \frac{\rho_g d_d U_{rel}}{\mu_g}$ | 200 | 1996 | 454 | 10479 | 984 | 71537 | 191 | 3373 | 1018 | 3074 | 1 | 22954 |
| Capillary number | $Ca = \frac{\mu d_d U_{rel}}{\sigma d_p}$ | $1.10^{-4}$ | $5.10^{-3}$ | $2.6.10^{-4}$ | $1.5.10^{-3}$ | $2.4.10^{-3}$ | $1.8.10^{-3}$ | $1.2.10^{-5}$ | $3.4.10^{-4}$ | $1.3.10^{-4}$ | $1.5.10^{-3}$ | $4.5.10^{-5}$ | 1000 |
| Ohnesorge number | $Oh = \frac{\mu}{\sqrt{\sigma \rho_d d_d}}$ | $5.2.10^{-3}$ | 0.012 | $4.9.10^{-3}$ | $6.2.10^{-3}$ | $1.6.10^{-3}$ | $2.2.10^{-3}$ | $2.2.10^{-3}$ | | 0.007 | 0.009 | 0.013 | 2.2 |
| Laplace number | $Lp = \frac{\sigma \rho_d d_d}{\mu^2}$ | 7300 | $3.6.10^{4}$ | 260 | $4.2.10^{4}$ | 2100 | $4.1.10^{5}$ | $2.1.10^{5}$ | | $1.2.10^{4}$ | $2.10^{4}$ | 0.2 | 6400 |
| Impact parameter | $b = \frac{2B}{(d_d + d_p)}$ | 0 | 1 | 0 | 1 | 0 | sin(60º) | 0 | 1 | 0 | | 0 | 1 |
| d:p diameter ratio | $\Delta = \frac{d_p}{d_d}$ | 0.5 | 1 | 3.05 | 12.31 | 0.18 | 0.34 | 0.85 | 1.36 | 1.79 | 11.76 | $1.10^{-5}$ | 10 |
| Stokes number | $St_k = \frac{U_{rel} \rho_d d_d^2}{d_p 9 \mu_g}$ | $1.2.10^{3}$ | $6.1.10^{4}$ | $5.7.10^{3}$ | $2.1.10^{4}$ | $1.8.10^{5}$ | $2.5.10^{6}$ | $1.4.10^{3}$ | $1.5.10^{4}$ | $2.1.10^{3}$ | $1.1.10^{4}$ | $6.1.10^{2}$ | $1.4.10^{10}$ |
| Deposition efficiency | $DE = \frac{St_k^2}{(St_k + 0.25)^2}$ | 1 | 1 | 1 | 1 | 1 | 1 | 1 | 1 | 1 | 1 | 0.999 | 1 |

The densities of these components range from 1100 to 13500 kg.m$^{-3}$ and they present wettability characteristics from hydrophilic to hydrophobic.

The interactions of heterogeneous droplets and particles with the wide range of physico-chemical properties discussed above necessarily means that the system will encompass hydrodynamic regimes where different principles apply. As already noted above, these systems will move from the Stokes to the Newton Reynold's number regimes and will traverse the limits identified by the Ohnesorge and capillary number thresholds described previously. An assessment must therefore be made of whether there are particular regions of the droplet-particle interaction map that are the most critical to study in order to ascertain the impact of collisions on the transmissibility of viruses contained in the droplets, and to determine whether the model chosen here will be suitable. A simple starting point for this assessment is to consider the overall probability of collision in a well-mixed volume with populations of droplets and particles possessing the size distribution characteristics described above. The probability of collision, $p_{coll}$, in this space for a particular droplet-particle pair is therefore:

$$p_{coll} = n_d n_p \pi \left(r_d + r_p\right)^2 \tag{12}$$

Where $n_{d,p}$ is the number of droplets or particles and $r_{d,p}$ is the radius of droplet or particle. This simple relation assumes unitary velocity and volume and neglects the nature of the introduction of the droplets and particles into the environment: in reality aerosol droplets are introduced into the air through breathing, talking, coughing and sneezing, each of which can be modelled as an interrupted jet [27, 29, 33] with different starting characteristics with respect to droplet concentration, time, volume and velocity. Three initial velocities are considered in this work: for breathing or speaking, a velocity of 3 m.s$^{-1}$ is used, following measurements by Chao et al, Kwon et al and Tang et al [34-36]. Similarly, coughs have been studied extensively and velocities in the range of 1.5 to 28.8 m.s$^{-1}$ have been recorded [34-37]. Here we will use a cough velocity of 11 m.s$^{-1}$ as the mean of the mean velocities measured in these works. Sneezes [27, 36, 38] have been less susceptible to quantitative analysis and precise velocities have not been reported, beyond an understanding that they are propelled at greater velocities than coughs. The exception to both of these statements is the work of Tang et al [36] who have measured the velocities of sneezes as surprisingly not exceeding 4.5 m.s$^{-1}$. We have used a velocity of 23 m.s$^{-1}$ for sneezes in the present study, as a mid-point between the measurements of Tang et al and the older estimates of anywhere from 20 to 50 and even 100 m.s$^{-1}$. This high value also represents the upper end of the velocity range for coughs. A thorough analysis of the impact of this injection of droplets into the dilute solution of particles and the subsequent evolution of the velocity of collisions will be reported elsewhere. In the following we will neglect collisions between droplets and between particles.

A heat map of the collision probability $p_{coll}$ for the distributions of droplets and particles is shown in Figure 4a. Probabilities span 6 orders of magnitude, with the highest probability found for the largest droplets with the smallest particles and the lowest probability where the smallest droplets encounter the largest particles. This latter regime where larger particles collide with smaller droplets, or $p > d$, represents $1\times10^{-7}$ of the total collisions and is labelled in black in figure 4a. The evolution of Reynolds number and Ohnesorge number against droplet size can be scaled to the collision probability for each droplet and the thresholds between Newtonian and intermediate flow regimes, and where dissipative energy processes become important, are shown in figures 4b and c. It can be seen from this that the overwhelming majority of collisions take place in regimes ($Re > 200$ and $Oh < 0.115$) where the model of Ashgriz and Poo [8] was found to fit experimental data very well. Only the lowest velocity/highest viscosity case falls entirely into the intermediate Reynolds number regime and at the corresponding We numbers collisions in this region would be expected to result in coalescence and not separation. These results establish the validity of an attempt to adapt the droplet-droplet models of Ashgriz and Poo [8] and Ko and Ryou [9] to the droplet-particle case, and in particular to respiratory aerosol droplet-PM particle collisions, to generate We-b maps and predict the fate of liquid in the initial droplet post-collision.

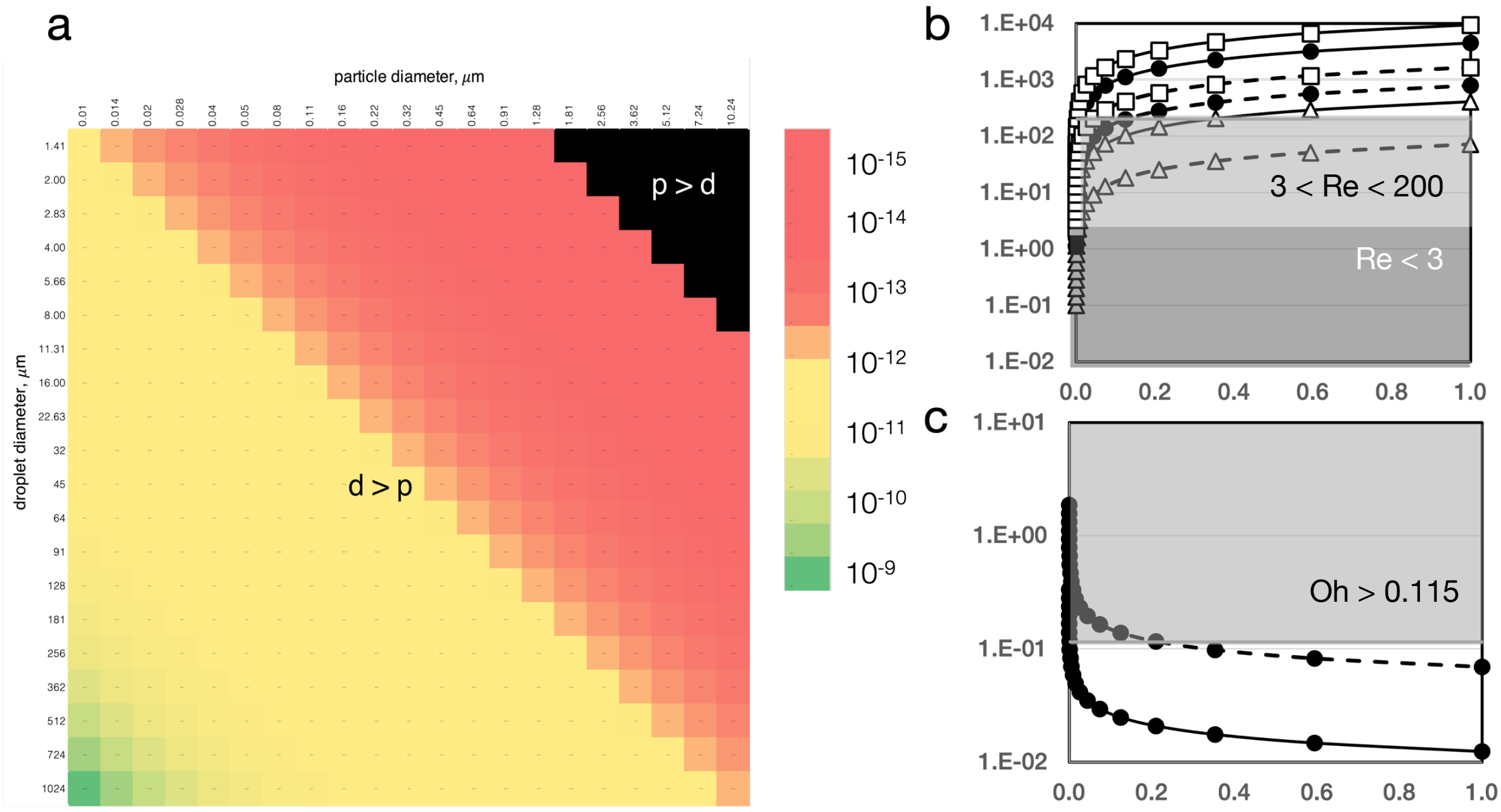

**Figure 4:** (a) Heat map of the collision probabilities for each aerosol droplet: PM particle pair. The region in black at top right is where the particle is larger than the droplet in the interaction. This case applies for $1\times10^{-7}$ of collisions. (b) Plot of Reynolds number Re plotted against cumulative probability of collision for each droplet size, from smallest to largest. Dashed lines are for liquid viscosity of $14.2\times10^{-2}$ mPa.s, solid lines are for a viscosity of $2.5\times10^{-3}$ mPa.s, open triangles represent $U_{rel} = 3$ m.s$^{-1}$, filled circles $U_{rel} = 11$ m.s$^{-1}$ and open squares $U_{rel} = 23$ m.s$^{-1}$. The two shaded regions at Re = 3-200 and at Re > 3 represent the boundaries between Newtonian, intermediate and Stokes behaviour. For all cases except the lowest velocity/highest viscosity case, the Newtonian Reynolds regime applies for all or a majority of collisions. (c) Plot of Ohnesorge number against cumulative collision probability for each droplet size. Dashed line represents $\mu_d = 14.2\times10^{-2}$ mPa.s, solid line represents $\mu_d = 2.5\times10^{-3}$ mPa.s. The shaded region represents Oh > 0.115. More than 70% of high viscosity collisions, and essentially all the low viscosity collisions, take place in the low Oh region where dissipative processes do not affect the model fits.

### 2.4.1 Reflexive separation in particle-droplet collisions

Following Ashgriz and Poo [8], our starting point remains a consideration of the energy flows within the droplet and particle. In order to determine the outcomes of collisions between droplet and particles, we need to re-examine the sources of each of these energy flows. As figure 5(a) shows, the excess surface-induced energy $KE_{ES}$ arises from the difference in surface energy of the parent droplets and that of the combined mass. When the smaller droplet $d_1$ is substituted for a solid particle p with interfacial tensions with air and droplet $\sigma_p$ and $\sigma_{pd}$, contact angle $\theta$, density $\rho_p$, volume $V_1$ and velocity $\vec{u}_1$, the surface energy terms change as the remaining droplet $d_2$ creates a new interface with the surface of the particle while the particle loses an interface with the air. The size of this change will be determined by the contact angle of the droplet medium on the particle surface according to a rearrangement of Young's equation, $\sigma_{pd} - \sigma_p = -\sigma\cos(\theta)$ so that $KE_{ESp}$ (eq. S1) is rewritten as:

$$KE_{ESp} = \sigma\pi d_2^2\left[(1 - \Delta^2\cos\theta) - (1 + \Delta^3)^{2/3}\right] \tag{13}$$

The relative magnitudes of eqs. (S1) and (13) will depend upon the experimentally measured value of the contact angle of the droplet medium on the particle surface.

Similarly, the counteractive and stretching flow terms given by eqs. (S2) and (S3) contain the interacting volumes $V_{1i}$ and $V_{2i}$, which represent the fractions of the droplets that deform and experience counteractive flow. For a non-deformable particle the corresponding term $V_{1i}$ will be zero, so that the result of subtracting $KE_{ST}$ from $KE_{CO}$ will be:

$$KE_{CO} - KE_{ST} = \frac{\rho}{2}[(2V_{2i} - V_2)\vec{u}_2^2] - \frac{\rho_p}{2}V_1\vec{u}_1^2 \tag{14}$$

The resulting term for $KE_{Rdp}$ is then given as:

$$KE_{Rdp} = \sigma\pi d_2^2\left[(1 - \Delta^2\cos\theta) - (1 + \Delta^3)^{2/3}\right] + \frac{\rho}{2}[(2V_{2i} - V_2)\vec{u}_2^2] + \frac{\rho_p}{2}V_1\vec{u}_1^2 \tag{15}$$

Rewriting eq. (15) using the substitutions for $\eta_1$, $V_{2i}$, $\vec{u}_1^2$ and $\vec{u}_2^2$ described in eqs. (S4-8), and defining a new density ratio parameter $P = \rho_p/\rho_d$, we find:

$$KE_{Rdp} = \sigma\pi d_2^2\left[(1 - \Delta^2\cos\theta) - (1 + \Delta^3)^{2/3}\right] + \frac{\rho\pi d_2^3\vec{u}^2}{12}\frac{\Delta^3}{(1 + \Delta^3)^2}(\Delta^3\eta_1 + \mathrm{P}) \tag{16}$$

Applying the boundary condition for reflexive separation given in eq. (S14) we have:

$$We_{Rdp} = 3\Delta(1 + \Delta^3)^2\left[\frac{7(1 + \Delta^3)^{2/3} - 4(1 - \Delta^2\cos\theta)}{(\Delta^6\eta_1 + \Delta^3 P)}\right] \tag{17}$$

Figure 5(b) shows the data collected by Pawar et al for $\Delta = 0.85$ (reflecting the case where the particle is smaller than the droplet), with the solutions for We given by Ashgriz and Poo for droplet-droplet collisions (eq.15) and that derived here for droplet-particle collisions using $P = 2.53$ and $\theta = 68º$ (eq. 17) [15]. Equilibrium contact angles specified in [15] ranged between 70º and 80º, but due to contact angle hysteresis the relevant contact angle is the receding contact angle. Contact angle hysteresis (advancing contact angle $\theta_a$ – receding contact angle $\theta_r$) for water on glass has been estimated as ±5.5-8.5º [39] so a value for the effective receding contact angle of 68º is assumed in this work. The regime plot shows that the reflexive separation boundaries for both droplet-droplet and droplet-particle collisions occur at values of We higher than was accessed experimentally, and this then explains why Pawar et al did not observe reflexive separation in their experiments. It is notable however that the d-p reflexive separation boundary extends much further into the b-We space, and that therefore reflexive separation after the droplet-particle collision here is more likely than following a droplet-droplet collision. Figure 5(c) shows how the droplet-particle collision reflexive separation boundary changes with contact angle $\theta$ between 0º and 120º. This shows that hydrophilic particles with low contact angles favour coalescence of the (aqueous) droplet on the particle while the droplet separates more readily from the hydrophobic particle at all values of the impact parameter. Likewise figure 5(d) shows how changing the density of the particle relative to the droplet from 0.5 to 3 moves the reflexive separation boundary by increasing the kinetic energy of the particle for denser particles, favouring ripping or skirting outcomes over coalescence. Finally figure 5(e) shows how We varies with the particle-droplet size ratio $\Delta$ between 0.4 and 1. Here the dependence of We on $\Delta$ is complex, showing a maximum separation rate at $\Delta \approx 0.8$. For $\Delta < 0.4$ separation occurs at We values $\gg 100$ and there is no longer any dependence on b. Indeed for all the parameters discussed here the dependence on b is much weaker than for the droplet-droplet interaction. This arises from the loss of one of the terms containing b ($\eta_2$) in the denominator of the We boundary equation (cf. eqs. (S11) and (17)).

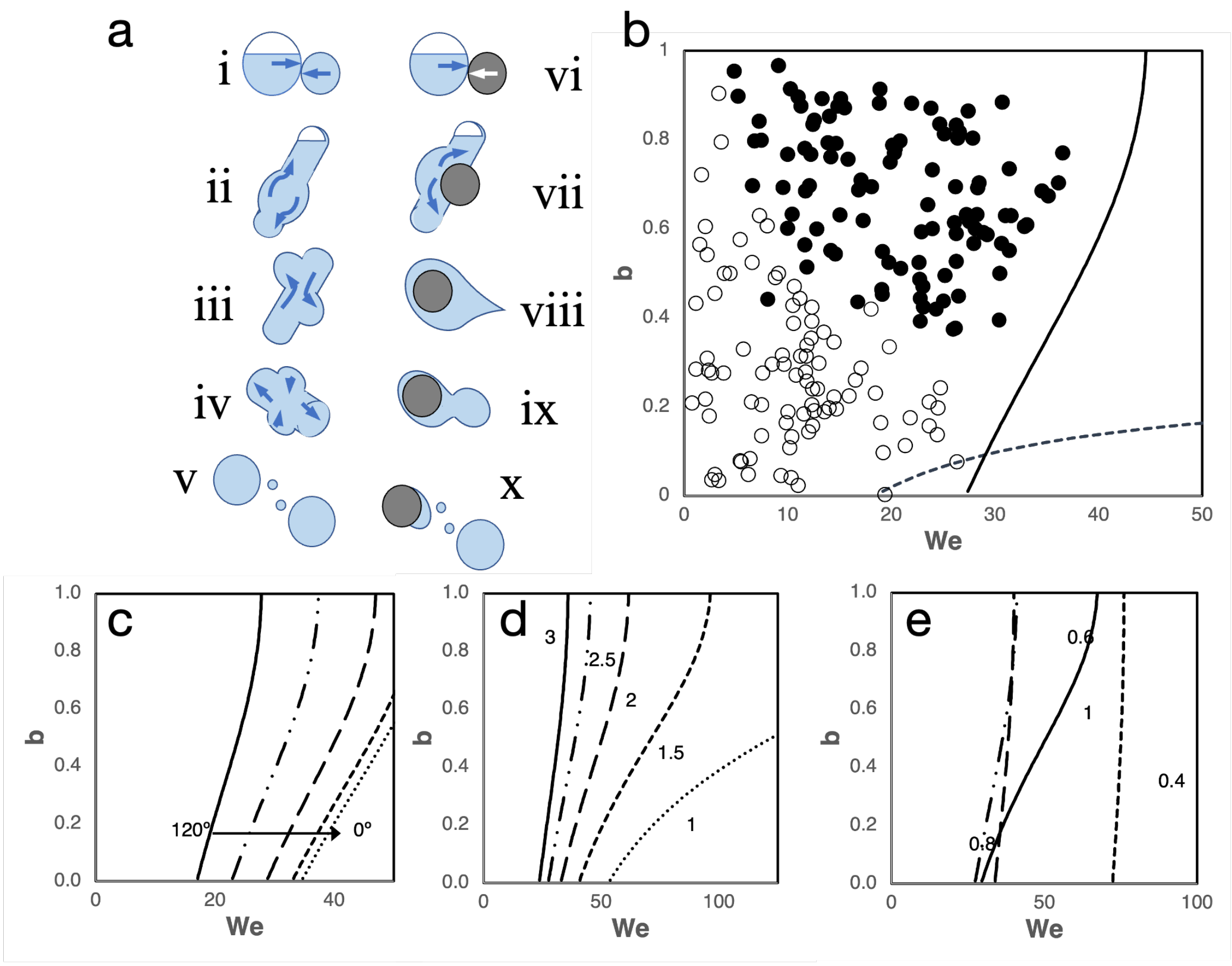

**Figure 5:** (a) illustration of the origin of the excess-surface-induced energy in droplet-droplet and droplet-particle collisions. At point i the droplets collide close to head-on. The shaded parts of the droplets are those parts which now begin to mix. At ii the droplets are mixing and form an extended oblate shape before contracting (iii and iv) to reduce surface area before separating (v) with the potential creation of satellite droplets. For particle-droplet collisions (vi) the process differs as the particle cannot share its material with the droplet, so the droplet deforms (vii) to engulf the particle (viii) before separating (ix), with the potential to create satellite droplets and to leave a residual volume of liquid on the particle (x). (b) We-b regime plot for the data collected by Pawar et al for the case where $\Delta = 0.85$. Filled circles represent separation collisions and open circles represent coalescence following collision. The dashed line shows the result of eq. (S11), the reflexive separation We boundary for droplet-droplet interactions, and the solid line shows the result of eq. (17), the reflexive separation We boundary for droplet-particle interactions derived in this work. (c) We-b regime plot showing the effect on the droplet-particle separation boundary of varying the contact angle of water with the particle surface from 0º to 120º. (d) We-b plot showing the effect of varying the relative density $P$ of the particle material between 1 and 3. (e) We-b plot showing the effect of varying the droplet-particle size ration *Δ* from 0.4 to 1. Note that for figures (d) and (e) the x-axes have been extended beyond the experimental data in order to show the separation boundaries.

### 2.3.2 Stretching separation in particle-droplet collisions

A similar consideration of the interacting volumes in a stretching separation shows that the interacting volume of the non-deformable particle $V_{1i}$ must be zero, so that the kinetic energies of the non-interacting parts of the particle and droplet are given by:

$$KE_{Sn} = \frac{P\rho}{2}(V_1)\vec{u}_1^2 + \frac{\rho}{2}(V_2 - V_{2i})\vec{u}_2^2 \tag{18}$$

and the kinetic energy of the interacting part of the liquid droplet is:

$$KE_{Si} = \frac{\rho}{2}V_{2i}(\vec{u}_2 b)^2 \tag{19}$$

so that the sum of the kinetic energies $KE_S = KE_{Sn} + KE_{SI}$ can be written as:

$$KE_S = \frac{P\rho}{2}\vec{u}_1^2 V_1 + \frac{\rho}{2}\vec{u}_2^2 V_2 + \frac{\rho}{2}\vec{u}_2^2 V_{2i}(b^2 - 1) \tag{20}$$

The surface energy acting against the stretching kinetic energy is estimated as that of a cylinder with volume $V_{2i}$, diameter $d_i$ and height h (see eqs. S13-15) with the difference now being that one end of the cylinder is creating a new interface with the surface of the particle (see figure 6(a)) so that:

$$SE_I = \sigma\left[(2\pi V_2 d_i \phi_2 \tau)^{1/2} + \frac{\pi\cos\theta}{2}\phi_2 d_i^2\right] \tag{21}$$

with:

$$d_i = 2d_2\sqrt{\frac{\phi_2}{3\tau}} \tag{22}$$

Substituting eqs. (S6), (S13-15) and (22) into eqs. (20) and (21) then gives:

$$KE_S = \frac{\rho d_2^3 \vec{u}^2 \Delta^3}{12(1+\Delta^3)^2}\left[\mathrm{P} - \Delta^3\left(1 - \phi_2(b^2 - 1)\right)\right] \tag{23}$$

$$SE_i = \sigma\pi d_2^2\left[\sqrt{\frac{2\phi_2\sqrt{\tau}}{3\sqrt{3}}} + \frac{2\,\phi_2\cos\theta}{3\tau}\right] \tag{24}$$

So that the stretching separation boundary condition for droplet-particle collisions expressed in dimensionless terms is:

$$We_{Sdp} = \frac{12(1+\Delta^3)^2\left[\sqrt{\frac{2\phi_2\sqrt{\tau}}{3\sqrt{3}}+\frac{2\,\phi_2\cos\theta}{3\tau}}\right]}{\Delta^2\left[P-\Delta^3\left(1+\phi_2(1-b^2)\right)\right]} \tag{25}$$

Figure 6(b) shows the We-b regime plot for Pawar et al's data for $\Delta = 0.85$, $P = 2.53$ and $\theta = 68^{\circ}$ with the droplet-droplet (dashed line, eq. (6)) and droplet-particle (solid line, eq. (25)) solutions for the stretching separation boundaries. The newly derived droplet-particle boundary fits the data very well and, again, requires only the two new experimentally measurable parameters $P$ and $\theta$. The effect of varying these two parameters can be seen in figures 6(c) and (d), for values of $\theta$ from 0º to 120º and values of $P$ from 1 to 3. The stretching separation is more highly favoured for hydrophobic particles ($\theta = 120^{\circ}$) and for denser particles. Finally the effect of varying the particle to droplet ratio $\Delta$ is shown in figure 5(e). Here the dependence of the separation boundary is more complex, showing a maximum separation rate at $\Delta = 0.5$ with minima at the extremes. As for the reflexive separation, the dependence of the We boundary on b is weaker than for the droplet-droplet case, due to the loss of the $\phi_1$ term from the denominator of eq. (25). And again it is clear that the droplet-particle collision is more likely to result in separation than the droplet-droplet collision.

It should be noted that, because both the reflexive and the stretching separations can be found to occur at any value of the impact parameter, there are regions at high values of We in the We-b regime plot where both types of separations are predicted to occur. In these cases, following Munnannur and Reitz [12] we can assume that the separation mode with the highest energy (i.e. the higher of eqs. (16) and (23)) is the mode that occurs.

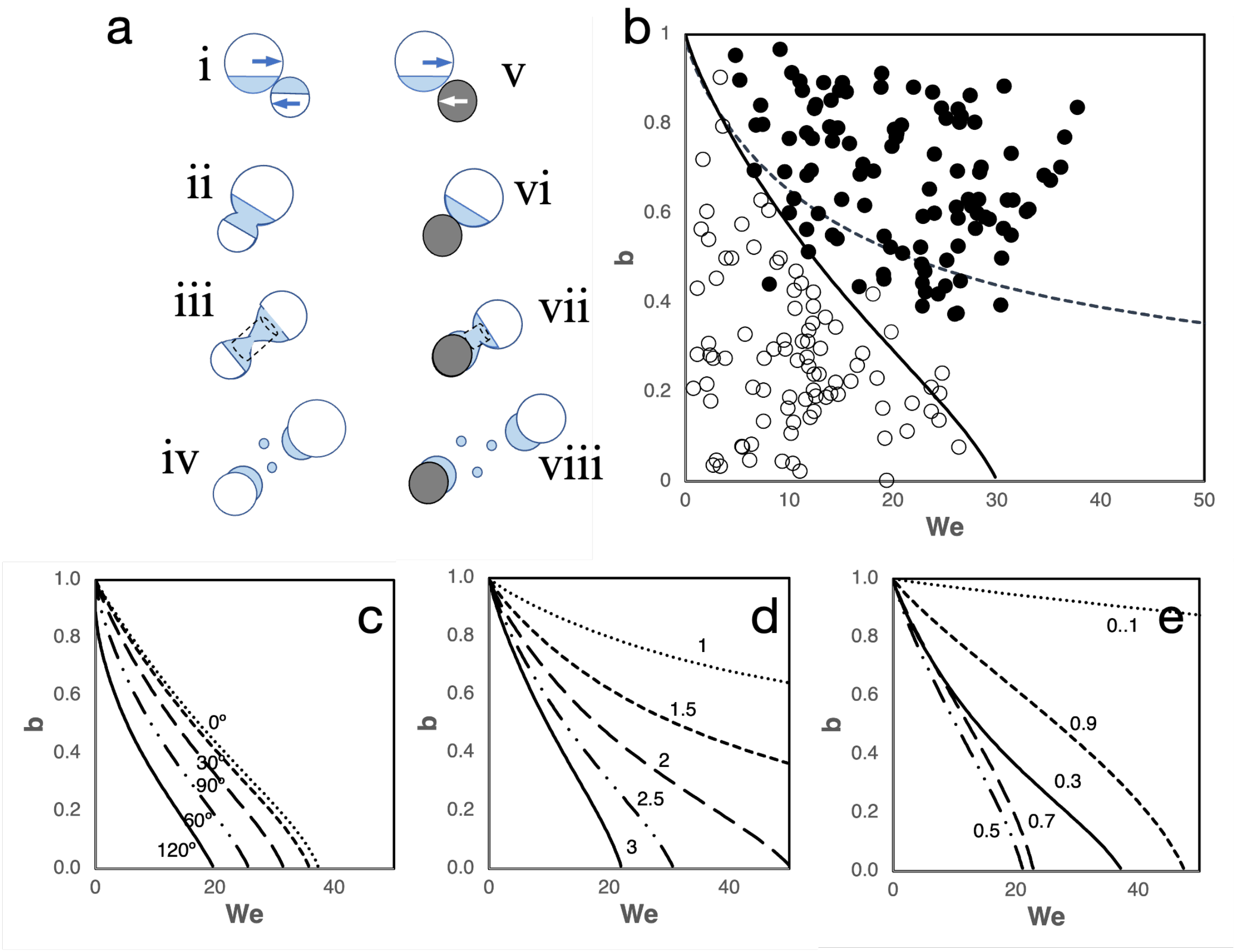

**Figure 6:** (a) illustration of the origin of the excess surface induced energy in droplet-droplet and droplet-particle collisions. Two droplets collide (i) and, because of the limited shared area between them only a fraction of their volume $V_i$ is shared (shown as shaded) (ii). As the droplets begin to separate (iii), this interacting volume can be approximated as a cylinder with diameter $d_i$ and volume $V_i$ and this volume is then, following separation (iv), distributed between the leaving droplets and any satellite droplets. When one of the droplets is substituted for a particle (v), all of the interacting volume must come from the droplet (vi), causing greater deformation of the droplet to minimise surface area. The diameter of the interacting cylinder (vii) now depends on the interfacial energy of the particle-droplet interface and on separation (viii), the volume is shared between the leaving droplet, any satellite droplets and any residual volume left on the particle.. (b) We-b regime plot for the data collected by Pawar et al for the case where $\Delta = 0.85$. Filled circles represent separation collisions and open circles represent coalescence following collision. The dashed line shows the result of eq. (6), the We stretching separation boundary for droplet-droplet interactions, and the solid line shows the result of eq. (25), the We stretching separation boundary for droplet-particle interactions derived in this work. (c) We-b regime plot showing the effect on the droplet-particle separation boundary of varying the contact angle of water with the particle surface from 0º to 120º. (d) We-b plot showing the effect of varying the relative density $P$ of the particle material between 1 and 3. (e) We-b plot showing the effect of varying $\Delta$.

**2.3.3 post-stretching separation characteristics in particle-droplet collisions**

The outcomes of collisions predicted by the new droplet-particle collision models that were illustrated in figures 5 and 6 can be used as inputs to Ko and Ryou's model (eqs. 7-11) to predict the size, number and velocities of the droplets post-collision. Following a separation event there will be a number of participating entities: Besides the original particle there may be an agglomerating fraction of the droplet that remains adsorbed on to the particle, there will be the departing droplet and there may be a number of satellite droplets. Proceeding from Ko and Ryou's arguments for the post-stretching separation characteristics, the mass and momentum balance equations for these entities can be written as:

$$Pd_1^3 + d_2^3 = Pd_1^3 + \Psi d_2^3 + N_{sa}d_{sa}^3 + (1 - C_{VS}\phi_2)d_2^3 \tag{26}$$

$$Pd_1^3\vec{u}_1 + d_2^3\vec{u}_2 = Pd_1^3\vec{u}_{1a} + \Psi d_2^3\vec{u}_{1a} + N_{sa}d_{sa}^3\vec{u}_{sa} + (1 - C_{VS}\phi_2)d_2^3\vec{u}_{2a} \tag{27}$$

where we introduce a new non-dimensional term $\Psi$ for the volume fraction of the droplets that remains as a residue on the particle post-separation. This residual volume, alongside the volume of any satellite droplets, comes from the volume fraction $\phi_x C_{VS}$ so that:

$$C_{VS}\phi_2 d_2^3 = \Psi d_2^3 + N_{sa}d_{sa}^3 \tag{28}$$

From eq. (28) we can obtain an expression for $d_{sa}$, the diameter of a satellite droplet, although we do not yet have an expression for $N_{sa}$:

$$d_{sa} = \sqrt[3]{\frac{C_{VS}\phi_2 - \Psi}{N_{sa}}}\, d_2 \tag{29}$$

and by assuming that the velocities of the droplet and particle are only changed post-collision by dissipative energy loss [9], we can use the conservation of momentum to determine an expression for $\vec{u}_{sa}$:

$$\vec{u}_{sa} = \frac{-\vec{u}_{rel}}{(1+\Delta^3)} \frac{d_2^3}{d_{sa}^3} \frac{(\Psi + \Delta^3 C_{VS}\phi_2)}{N_{sa}} \tag{30}$$

In order to determine the value of Ψ, we now assume that the shape the droplet adopts on the surface of the (also assumed) spherical particle is itself a segment of a sphere, which images of droplets on spherical particles [40] show to be a reasonable approximation. The value of Ψ can be determined if we assume that the volume $V_r$ equals that of the volume of a droplet with radius $R_d$, minus the volume of the overlapping segment of the sphere with the spherical particle as depicted in figure 7, and given by:

$$V_r = \frac{4}{3}\pi R_d^3 - \frac{\pi}{3}\left(R_p^3 + R_d^3\right)(2 + \cos\theta_s)(1 - \cos\theta_s)^2 \tag{31}$$

where $\theta_s$ is the spreading angle defined below, $R_p$ is the radius of the particle ($= \Delta d_2/2$) and $R_d$ is the apparent radius of the droplet deposited on the particle. Figure 7 shows the overlapping segment as the superposition of the two spheres and defines the distances and angles that need to be found to quantify the volume and the areas of the two surfaces (droplet-air and droplet-particle). Extrand and Moon [40] used Gibbs' relation for describing how the contact angle is influenced by a sharp edge [41] to derive a simple relation between the apparent (advancing or receding) contact angle measured on the spherical surface $\theta_{(a,r)}$ and the intrinsic (advancing or receding) contact angle $\theta_{(a,r)o}$ that represents what the contact angle would be on a flat surface, with the difference being defined as the spreading angle $\theta_s$ as given below:

$$\theta_s = \theta_{(a,r)} - \theta_{(a,r)o} = \sin^{-1}\left(\frac{c}{R_p}\right) \tag{32}$$

where c is the chord that intersects the contact lines on either side of the drop, passing through the centre axis of the deposited drop, and the subscript (a,r) represents advancing or receding contact angle. As we are modelling the process of separation, we will use the apparent and intrinsic *receding* contact angles. $R_d$ is found from the quantity $x$ (see Figure 7) and the length of the chord c is initially found from eq. (22) since $c = d_i/2$ and so the instantaneous spreading angle $\theta_s'$ and the instantaneous apparent receding contact angle $\theta_r'$ are determined from the interaction cylinder and depend on the impact parameter. For small values of the impact parameter, $\theta_s'$ calculated this way is larger than the spreading angle determined from equilibrium measurements of $\theta_r$ and $\theta_{r,o}$, i.e. the droplet wets a larger surface area of the particle than it would at equilibrium, Thus the contact area reduces as the droplet passes the particle until an equilibrium contact angle defined by a spreading angle of $\theta_s$ is reached. We can then find $\theta_s'$ and, for a material with a known $\theta_{r,o}$, $\theta_r'$, from eq. (32), and $R_d$ from:

$$x = c\tan(90^{\circ} - \theta_r),$$

$$R_d = \sqrt[2]{c^2 + x^2} = \frac{\Delta d_2}{2}\sec(90^\circ - \theta_r)\sqrt[2]{\sin(min\theta_s, \theta_s')} \tag{33}$$

And, from eq. (26), Ψ is simply:

$$\Psi = \frac{6V_r}{\pi d_2^3} \tag{34}$$

So that if we know the radius of the particle and the apparent and intrinsic receding contact angles of the droplet medium on the particle, we can determine Ψ.

The number of satellite droplets depends on the kinetic and surface energies of the newly created interfaces and is determined from the conservation of energy [9]:

$$KE_1 + SE_1 + KE_2 + SE_2 = KE_{1a} + SE_{1a} + KE_{2a} + SE_{2a} + KE_s + SE_s + DE \tag{35}$$

$KE_1$, $KE_2$ and $KE_{2a}$ are:

$$KE_{n,a} = \frac{\rho_n \pi}{12} d_{n,a}^3 \vec{u}_{n,a}^2 \tag{36}$$

$SE_1$, $SE_2$ and $SE_{2a}$ are:

$$SE_{n,a} = \sigma_n \pi d_{n,a}^2 \tag{37}$$

where $d_{n,a}$ and $\vec{u}_{n,a}$ are given in eqs. (7a) and (8a), save that $d_{1a} = d_1$. The term for the dissipative energy loss DE arises from viscous dissipation by the volume of liquid in the interacting cylinder so that:

$$DE = \frac{\alpha}{2}[\rho_2(1 - C_{VS}\phi_2) d_2^3 \vec{u}_2^2] \tag{38}$$

where α is an energy loss coefficient empirically estimated to be 0.5 [11]. This leads to an expression for the velocities of the droplet and particle after collision: $\vec{u}_{na} = \vec{u}_n$ which is the same as eq. (8a). $KE_{1a}$ and $SE_{1a}$ include terms for the particle post-collision and the residue of the droplet that remains and can be expressed as:

$$KE_{1a} = \frac{\pi}{12}\rho d_2^3 (\mathrm{P}\Delta^3 + \Psi)\vec{u}_{1a}^2 \tag{39}$$

$$SE_{1a} = \sigma\pi\left[\cos\theta_{r,o}\frac{\Delta^2}{2}d_2^2\cos\theta_s + R_d^2(1-\cos\theta_s)\right] \tag{40}$$

$KE_s$ is then the kinetic energy of the material in the satellite droplet(s):

$$KE_S = \frac{\pi}{12}\rho d_2^3 (C_{VS}\phi_2 - \Psi)\vec{u}_{sa}^2 \tag{41}$$

and $SE_s$ can be determined from eqs. (35-41) so that $N_{sa}$ is then given by:

$$N_{sa} = \left[\frac{SE_s}{\sigma\pi d_2^2 (C_{VS}\phi_2 - \Psi)^{2/3}}\right]^3$$

$$ \tag{42} $$

So that the post-collision characteristics ($\vec{u}$, d, Ψ, N) of the particle, droplet and any satellite droplets are given by $d_{1a} = d_1$, eqs. (7a), (8a), (29), (30), (34) and (42) and can be determined when we know the values of Δ, We, b, P, $\theta_r$ and $\theta_{r,o}$ for the droplet and particle materials involved in the collision.

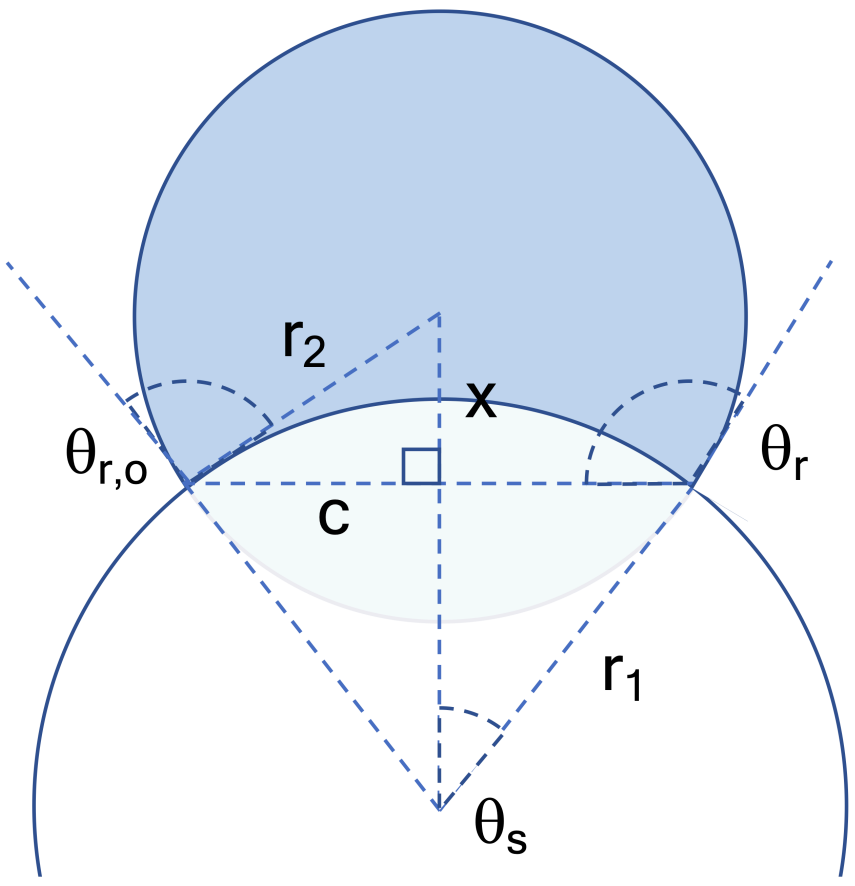

**Figure 7:** Depiction of the droplet (shaded) leaving the particle surface, showing the angles $\theta_s$, $\theta_r$ and $\theta_{r,o}$ and the distances $r_1$, $r_2$, c and x that are defined in the text.

**2.3.4 post-reflexive separation characteristics in particle-droplet collisions**

In reflexive separations many aspects of the process described above apply, so that the expressions for conservation of mass and momentum, and for $d_{sa}$ and $\vec{u}_{sa}$ (eqs. (26-29)) remain the same except for replacing the term $C_{VS}.\phi_2$ with the term $C_{VR,}$ and the expressions for the conservation of energy and for $KE_{1,\ 2}$ and $_{2a}$, and $SE_{1,\ 2}$ and $_{2a}$ (eqs. (35-37)) also remain the same. Where the post-collision characteristics differ is that whereas for stretching separation the contact interface between the particle and the droplet is confined to the end of the interaction cylinder with diameter $d_i$ (eq. (22)), in a reflexive separation essentially the whole particle is engulfed by the larger droplet during collision. Therefore, instead of starting from a small interaction area determined by the impact parameter, now the size of the remaining droplet on the particle will be determined solely by the properties of the droplet-particle interface. Thus the chord c is now determined from eq. (32) given our prior knowledge of $\theta_r$ and $\theta_{r,0.}$ This allows us to find $R_d$ and Ψ from eqs. (33) and (34), substituting $C_{VS}.\phi_2$ for $C_{VR}$ as above. The dissipative energy DE now arises from the viscous dissipation during the coalescence of the droplet on the particle and is:

$$ DE = \frac{\alpha}{2}[\rho_2 V_2 \vec{u}_2^2] \tag{43} $$

It is assumed, following eq. (8b), that $\vec{u}_{na} = \sqrt{2}/2\,\vec{u}_n$ and then $KE_{1a}$, $SE_{1a}$, $KE_s$ and $N_{sa}$ can be found from eqs. (39-42) using equations (26-38) with $C_{VR}$ instead of $C_{VS}.\phi_2$.

**2.3.5 Separation boundaries and characteristics for cases where p > d**

The preceding sections have all been concerned with the case where the particle is substituted for the smaller droplet, $d_1$. As was determined above, for the particular real interaction we wish to predict, this is the case for essentially all collisions. For the case where the particle is larger than the droplet, i.e. the particle is substituted for the larger droplet $d_2$, we can derive the equations below for the boundary thresholds for reflexive and stretching separation, in place of eqs. (17) and (25):

$$We_{Rdp} = 3\Delta(1+\Delta^3)^2 \left[\frac{7(1+\Delta^3)^{2/3} - 4(1-\Delta^2\cos\theta)}{\left(\eta_1 + 1 + \Delta^3(\Delta^3 P - 1)\right)}\right] \tag{44}$$

$$We_{Sdp} = \frac{12(1+\Delta^3)^2\left[\sqrt{\frac{2\phi_1\sqrt{\tau}}{3\sqrt{3}}} + \frac{2\,\phi_1\cos\theta}{3\tau}\right]}{\left(\Delta^6 P - \left(1+\phi_1(1-b^2)\right)\right)} \tag{45}$$

Figures 8a and b show the separation boundaries defined by eqs. (44) and (45) on plots of the experimental data [15] for the case of a particle of 4 mm and a droplet of 2.939 mm, showing that the adapted models successfully predict the stretching separation boundary, and that no reflexive separation occurs in this dataset. Figures 8c-e and f-g show how the stretching and separation boundaries change with varying material properties of the particle-droplet system, with the same dependencies seen as for the d>p case. Post-separation characteristics for stretching separation are then:

$$d_{1a} = (1 - C_{VS}\phi_1)^{1/3} d_1 \tag{46}$$

$$\vec{u}_{na} = \vec{u}_n \tag{47}$$

$$N_{sa} = \left[\frac{SE_s}{\sigma\pi\Delta^2 d_2^2 (C_{VS}\phi_1 - \Psi)^{2/3}}\right]^3 \tag{48}$$

$$d_{sa} = \left[\frac{C_{VS}\phi_1 - \Psi}{N_{sa}}\right]^{1/3} \Delta d_2 \tag{49}$$

$$\vec{u}_{sa} = \frac{\Delta^3 d_2^3}{(1+\Delta^3)}\vec{u}_{rel}\left[\frac{\Delta^3\Psi + C_{VS}\phi_1}{N_{sa} d_{sa}^3}\right] \tag{50}$$

And post-reflexive separation characteristics will follow the same pattern.

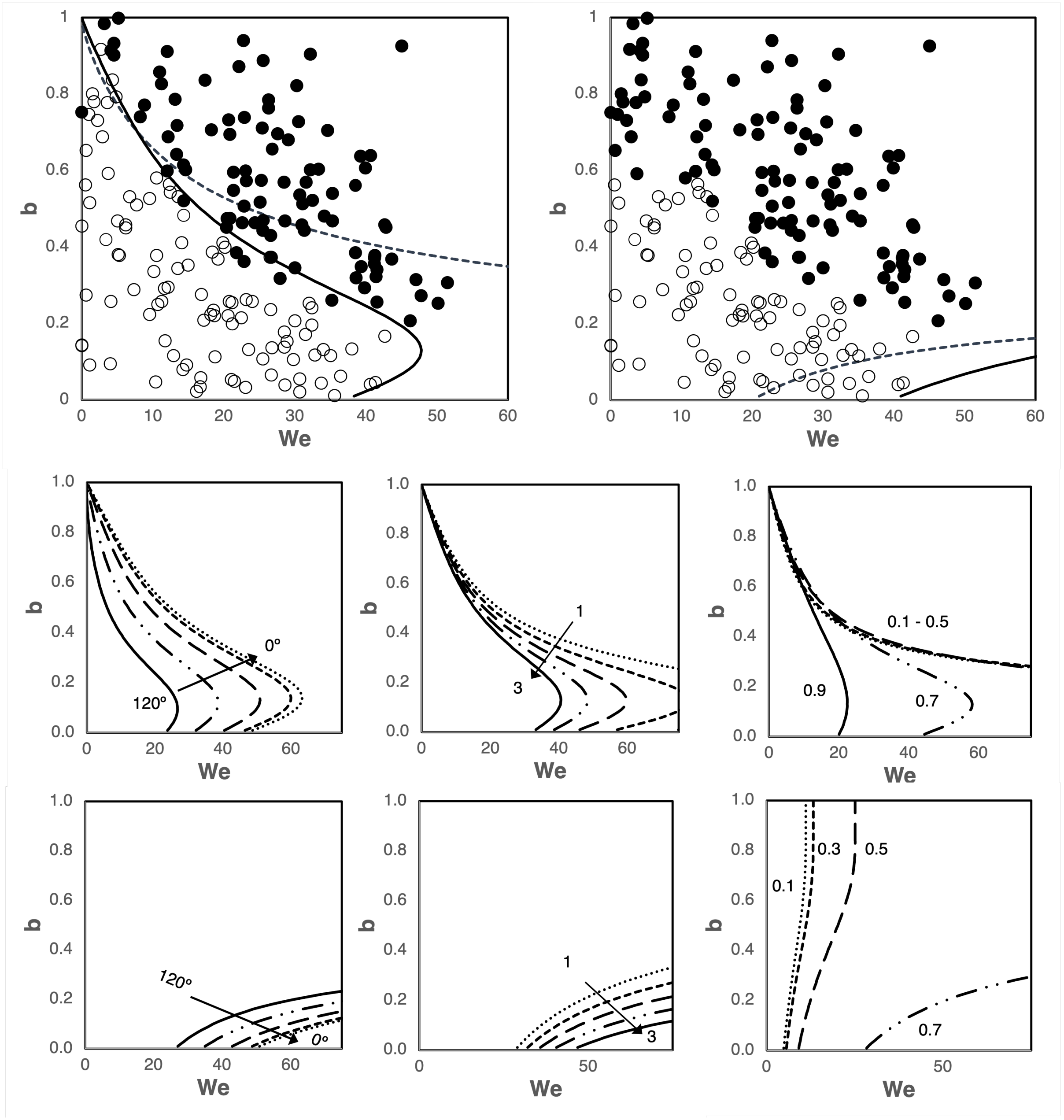

**Figure 8:** (a) We-b regime plot for the data collected by Pawar et al for the case where $\Delta = 1.36$. Filled circles represent separation collisions and open circles represent coalescence following collision. The dashed line shows the result of eq. (6), the stretching separation boundary for droplet-droplet interactions, and the solid line shows the result of eq. (45), the stretching separation boundary for droplet-particle interactions derived in this work. (b) as for (a) but dashed line shows the result of eq. (5), the stretching reflexive boundary for droplet-droplet interactions, and the solid line shows the result of eq. (44), the reflexive separation boundary for droplet-particle interactions derived in this work. (c, f) We-b regime plots showing the effect on the droplet-particle stretching (c) and reflexive (f) separation boundary of varying the contact angle of water with the particle surface from 0º to 120º. (d, g) We-b plots showing the effect of varying the relative density $P$ of the particle material between 1 and 3. (e, g) We-b plots showing the effect of varying $\Delta$.

## 3.1 Validation of the adapted model for post-separation characteristics of droplet-particle interactions

In contrast to the case for droplet-droplet collisions, experimental datasets for droplet-particle collisions and their outcomes across a wide range of the We-b space are scarce in the literature. Pawar et al [15] have provided a highly useful dataset for collisions between glass particles and water droplets with diameters in the mm range. Figures 5, 6 and 8 above have already shown how the model derived here successfully predicts the outcomes of collisions between 2.5 mm and 4 mm glass particles and 2.9 mm water droplets, revealing that at the range of We numbers attained in these experiments, coalescence and stretching separations, but not reflexive separations, can occur.

As no reflexive separations were observed in this dataset (or in recent work with similar properties [18]) we cannot validate the model for post-reflexive separation characteristics derived here against this experimental data, but we can compare the results of the model for the stretching separation to the observed data presented in figures 6-9 of [15]. Pawar et al calculated the agglomerated fraction left adhering to the particle following stretching separation from the initial masses and velocities of the particle and droplet and their velocities post-separation, as well as the velocities of any satellite droplets formed. From a momentum balance expression they then derived the extra mass adhering to the particle, neglecting the mass of any satellite droplets as these were observed to be very much smaller than the original droplet. Recently Islamova et al [18] repeated Pawar et al's experiment for $\Delta = 1.36$ and obtained very similar results. They only reported $E_{coal}$ and the number of satellite droplets formed.

**Table 2. Outcomes of stretching separation**

| | $V_r$ (%) | $V_{da}$ (%) | $N_{sa}$ | $V_{sa}$ (%) |
|---|---|---|---|---|
| [15] exp., $\Delta$ = 0.85 | 14.7 | 85.3[a] | 1.31 | n.d.[a] |
| model, $\Delta$ = 0.85 | 14.9 | 75.7 | 1.31 | 8.9 |
| [15] exp., $\Delta$ = 1.36 | 35.9 | 64.1[a] | 1.34 | n.d.[a] |
| [18] exp., $\Delta$ = 1.36 | n.d. | n.d. | 1.44 | n.d. |
| model, $\Delta$ = 1.36 | 40.8 | 49.3 | 0.81 | 9.9 |

[a] In their analysis of their experimental data, Pawar et al considered the volume of the separating droplets $V_s$ to be negligible and the volume of the leaving droplet $V_{da}$ has been inferred from that.

Table 2 compares the measurable or derivable outcomes (volume adhering to the particle $V_r$, volume of the leaving droplet $V_{da}$ and number and volume of satellite droplets $N_{sa}$ and $V_{sa}$) of stretching separations in the experimental dataset to the predictions made by the model presented here. In determining the fraction of the droplet that remains adhering to the particle, the model agrees very well with the data. The mean number of satellite droplets in the experimental and calculated data also agree very well for the d>p case and reasonably well for the d<p case. Contrary to Pawar et al's observation that satellite droplets represented negligible volumes, the model presented here predicts the volume of satellite droplets to be approximately 9-10% of the volume of the initial droplet, and the differences between the data and the model's predictions of the volume of the leaving droplet are attributable to these differences in estimates of the volume of the separating droplets.

To further explore how well the model predicts the experimental data, the post-separation characteristics of each event are binned by impact parameter and compared in Figure 9. Figures 9(a) and (b) show maps of the predicted and observed final states of the liquid droplet post-collision, as a function of the impact parameter for the present model and Pawar et al's dataset respectively. As stretching separations become possible at $b = 0.3$ for $d > p$, so more of the liquid droplet ends up in a leaving droplet or in satellite droplets, and a progressively smaller fraction remains adhered to the particle. This view of the data shows that the predicted agglomerated fraction reduces more gradually than is observed in the experiments, but the trend of the experimental data is faithfully reproduced. Figures 9(c) and (d) show how $N_{sa}$ evolves with impact parameter for the model (c) and the two experimental datasets (d). These compare experiment (figures 6 and 7 in [15] and figure 7b in [18]) and model: in both experiments the number of satellite droplets reaches a plateau at around $b = 0.4$, continuing until $b = 0.8$-$0.9$ before decreasing as b approaches 1, while the model predicts that numbers of satellite droplets rise as b rises, reaching a maximum as b nears 1. A comparison of figures 9(a) and (c) show that $V_{sa}$ does not follow the same pattern as $N_{sa}$, but peaks at around $b = 0.7$, which coincides instead with the peak in $N_{sa}$ observed experimentally. This suggests that at the highest values of b, where the model predicts an increased $N_{sa}$ in contrast to the experimental observations, the satellite droplets are much smaller. Indeed Figure 9(e) shows how the diameter of satellite droplets varies with impact parameter, reaching a minimum of 400 μm as b approaches 1. These smaller droplets may be more likely to be missed by the image capturing apparatus used in [15] and [18], which may resolve the disagreement between model and experiment seen in the $N_{sa}$ distributions above. Figure 9(c) also shows how the velocity of the satellite droplets reaches a maximum at $b = 0.6$. Finally figures 9(e) and (f) show the diameter of the leaving droplet following separation with good agreement between experiment and model.

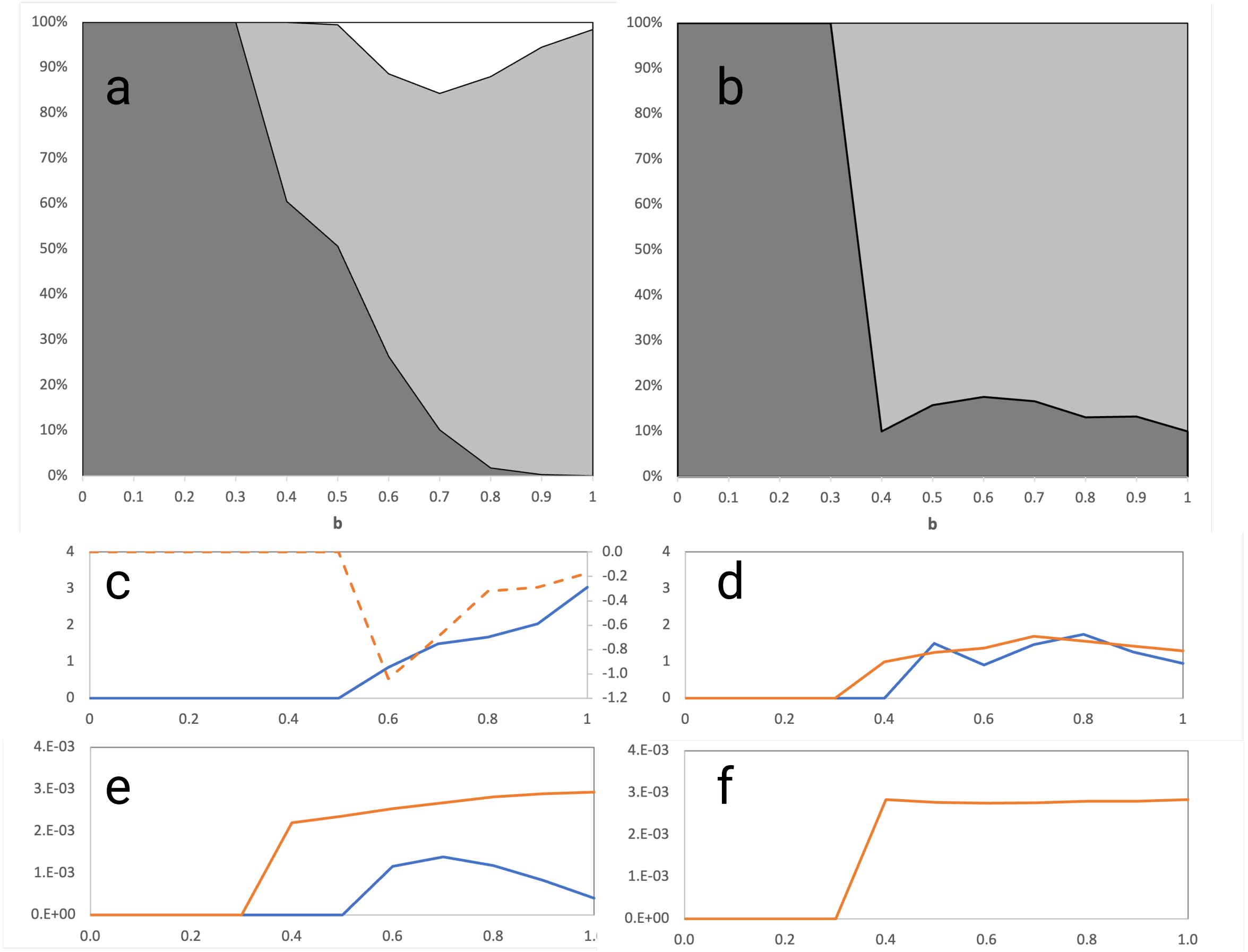

**Figure 9:** Plots of the volume fraction as a function of impact parameter of the collision for (a) the predictions of this model and (b) the results of Pawar et al [15] for $\Delta = 0.85$. Dark shading represents the droplet volume that remains on the particle, lighter shading represents the volume of the leaving droplet and white represents the volume of satellite droplets. (c) $N_{sa}$ (solid line, left-hand axis) and $u_{sa}$ (dashed line, right-hand axis) vs. impact parameter b for the simulation. (d) $N_{sa}$ for the experimental results of Pawar et al and for Islamova et al [18] vs. impact parameter b. (e) diameters of the leaving droplet (top) and satellite droplets (bottom) vs. impact parameter b for the simulation, (f) diameters of the leaving droplet vs. impact parameter b for Pawar et al.

The model can also be tested against the experimental observations of Dubrovsky et al, including data collected in high viscosity systems. This comparison may reveal the limits of the model in these conditions, which fall into the low Reynolds number and high Ohnesorge number regimes. Figure 10 reproduces figure 10 from Dubrovsky's work [13], showing the four collision outcomes identified by those authors on a Re-We map. The experiment was set up as described above, with head-on collisions, so b = 1. The two lines plotted on the figure represent two glycerol solution systems of differing viscosity: the high Reynolds number line (Re = 400-1500) reflecting a 30% solution and the low Reynolds number line (Re = 3-22) reflecting a 70% glycerol solution. The Ohnesorge values for the two systems are 0.054 and 4.24 respectively.

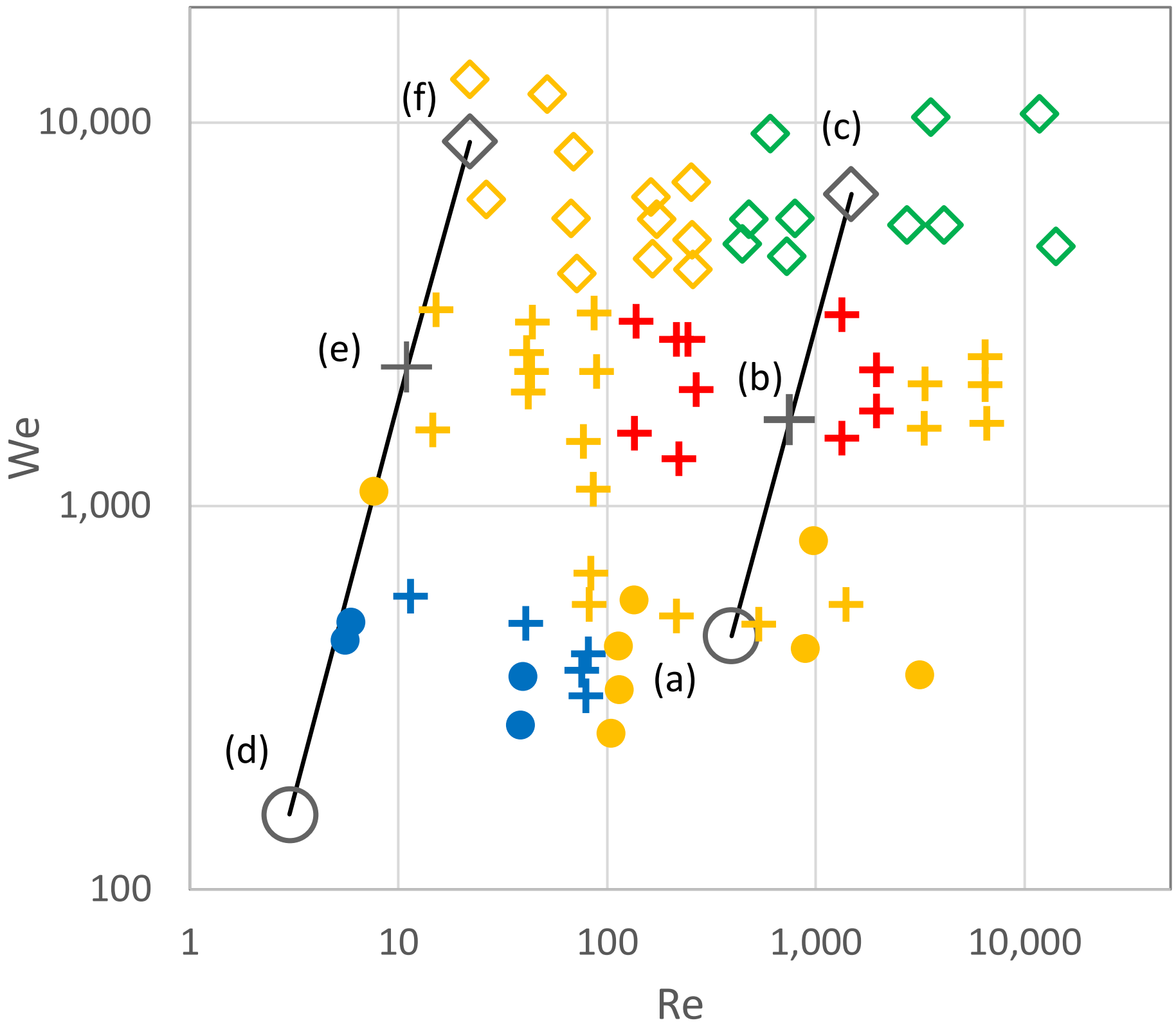

**Figure 10.** Results of collisions as recorded by Dubrovsky et al [13] plotted on axes of Weber number We and Reynolds number Re. Blue datapoints are 'capture' events, yellow datapoints are 'shooting-through' events, red points are 'destruction' collisions and green data are 'bubble' events, based on the terminology used by Dubrovsky et al. Filled circles are data collected at $U_{rel}$ = 3.4 m.s$^{-1}$, crosses are data collected at 6.5 m.s$^{-1}$ and open diamonds are data collected at 12.8 m.s$^{-1}$. The two lines linking black datapoints at the velocities described represent series of simulations using the parameters described in the text at these velocities. The outcomes of these simulations are presented in Table 3.

**Table 3.**

| Experimental outcome [13] | | We | Re | Oh | $U_{rel}$ $ms^{-1}$ | $V_r$ | $V_{da}$ | $V_{sa}$ | $d_{da}$ m | $N_{sa}$ | $d_{sa}$ m |
|---|---|---|---|---|---|---|---|---|---|---|---|
| shoot-through | (a) | 458 | 394 | 0.054 | 3.4 | 0% | 43% | 57% | $2.19 \times 10^{-3}$ | 1.68 | $2.02 \times 10^{-3}$ |
| destruction | (b) | 1675 | 750 | 0.054 | 6.5 | 0% | 14% | 86% | $1.49 \times 10^{-3}$ | 2.28 | $2.10 \times 10^{-3}$ |
| bubble | (c) | 6500 | 1480 | 0.054 | 12.8 | 0% | 4% | 96% | $9.63 \times 10^{-4}$ | 3.43 | $1.90 \times 10^{-3}$ |
| coalescence | (d) | 157 | 3 | 4.24 | 1.7 | 29% | 71% | 0% | $2.59 \times 10^{-3}$ | 0 | -- |
| shoot-through | (e) | 2300 | 11 | 4.24 | 6.5 | 0% | 14% | 86% | $1.50 \times 10^{-3}$ | 2.26 | $2.10 \times 10^{-3}$ |
| shoot-through | (f) | 8900 | 22 | 4.24 | 12.8 | 0% | 3% | 97% | $9.21 \times 10^{-4}$ | 3.57 | $1.88 \times 10^{-3}$ |

Table 3 presents the outcomes from the model for 3 points on each line (as indicated on Figure 10) along with the expected outcome based on Dubrovsky et al [13]. For the high Re line, the model predicts in all cases that no droplet material remains on the particle. In the lowest We case the leaving droplet contains 43% of the original droplet material with the rest forming satellite droplets. This outcome meets the description of a 'shooting through' event in which the particle passes through the droplet and the droplet remains intact, even if it has lost volume to satellite droplets. The position of this event on the Re-We map places it in the 'shooting through' region of Dubrovsky's data. The two higher We cases have 86 and 96% of the droplet material ending up in satellite droplets, with the original droplet left much smaller than the satellite volume. In this sense the droplet can be said to have been destroyed, and in Dubrovsky's map these two events fall in the 'destruction' and 'bubble' regions. The model therefore predicts similar behaviour in this system to the experimental observations, as expected in this high-Re, low-Oh, low viscosity regime.

The second system with low Re and high Oh gives similar predictions to the first – the two higher We cases have >86% of the droplet material leaving the collision in satellite droplets, with the initial droplet destroyed. The experimental observations in this area of the Re-We map suggest shooting-through events, where the droplet maintains its integrity. The datapoint at the lowest We value does show different behaviour, in that 29% of the droplet is predicted to agglomerate on the particle with the rest in a departing main droplet with no satellite droplets. A collision in this region of the Re-We map would be expected to result in coalescence rather than the shooting-through that the model predicts. An extension of the model to include viscosity and a dissipative energy component may be appropriate for systems with these properties.

### 4.1. Predicted outcomes of collisions between exhaled aerosol droplets and ambient pollutant particles

Finally, the model is applied to the aerosol:PM system in order to predict whether the size distribution of droplets post-collision is significantly changed from its initial distribution. 6 points are taken from the 2-d size distribution map in figure 4(a) (the map is divided into 6 areas of equal probability based on eq. (4) and droplet and particle diameters taken from the centre of each area) and their collisions at 3 different velocities (droplet velocities of 3, 11 and 23 $m.s^{-1}$ and mean particle velocity of 0) are simulated across the impact parameter range 0-1 for rare, water soluble particles (contact angle = 0º, $\rho = 1700$ $kg.m^{-3}$, modelling ammonium sulfate), medium density, hydrophobic particles (contact angle = 120º, $\rho = 2000$ $kg.m^{-3}$, modelling elemental carbon) and for dense, somewhat hydrophilic particles (contact angle = 60º, $\rho = 8000$ $kg.m^{-3}$, representing iron). These three particles represent common constituents of ambient PM in urban (carbon and iron) and rural (ammonium sulfate) environments.

The results of this simulation, as the fraction of the initial droplet volume that leaves the droplet during separation events, and the final distribution of the liquid from the original droplets among the agglomerated fraction, the leaving droplet and any satellite droplets, are presented in Table 4. Separations, resulting in smaller droplets leaving the particle, occur with all particles and velocities and overall occur in 12.4% of collisions. The outcomes are dependent on the droplet to particle diameter ratio $\Delta$ , on the density and contact angle of the particle and on the relative velocity $U_{rel}$, with collisions at high $U_{rel}$ and with particles with high $\Delta$ and $\rho$ much more likely to result in separations while contact angle is less influential in the final outcome of the collision. In all cases, stretching separations were observed rather than reflexive separations, even in head-on collisions, based on the kinetic energy criterion described above. The overall outcome across all 6 droplet collisions with three types of particle at three velocities is that 88% of the initial droplet volume is found in a coalesced droplet on the particle, 11% is in reduced droplets that leave the collision and 1% is in newly created satellite droplets. The diameters of leaving droplets following the collisions are between 1 and 71% of the initial droplet, while those of satellite droplets are between 0.001 and 36% of the initial droplet.

The current work presents a new model for the outcomes of collisions between droplets and particles, validates it against existing experimental data to establish the limits of its applicability and reports that the model predicts significant redistribution of droplet material between droplets of widely differing sizes, which may have consequences for the transmission of pathogens contained within them. To accurately assess the impact of collisions with ambient particulate matter on transmission of pathogens within respiratory droplets requires detailed knowledge of the evolution of the velocity, volume and concentration of droplets in the interrupted jet of the breath, cough or sneeze, of the frequency of these events and of the pathogenic concentration in each event, and of the relationship between the mass of the droplets and the time they remain airborne. This assessment will follow in a future report.

**Table 4. Outcomes of collisions between aerosol droplets and ambient PM**

| | | | | | | |
|---|---|---|---|---|---|---|
| d(d) | $1.02x10^{-3}$ | $1.41\ x10^{-6}$ | $5.67\ x10^{-6}$ | $3.20\ x10^{-5}$ | $1.02\ x10^{-3}$ | $2.56\ x10^{-5}$ |
| d(p) | $1.00\ x10^{-8}$ | $1.00\ x10^{-8}$ | $5.67\ x10^{-8}$ | $3.20\ x10^{-7}$ | $1.02\ x10^{-5}$ | $2.56E\ x10^{-6}$ |
| $\Delta$ | $9.77\ x10^{-6}$ | $7.09\ x10^{-3}$ | 0.01 | 0.01 | 0.01 | 0.1 |
| f(sep.) | 1.0% | 1.0% | 1.1% | 2.0% | 25.2% | 40.9% |
| vs. $U_{rel}$ | 3 | 11 | 23 | | | |
| f(sep.) | 1.8% | 10.2% | 23.6% | | | |
| vs. $\rho$ | 1700 | 2000 | 8000 | | % sep | 12.4% |
| f(sep.) | 6.0% | 11.8% | 17.8% | | $V_r$ | 88.1% |
| | | | | | $V_{da}$ | 11.2% |
| vs.$\theta$ | 0 | 60 | 120 | | $V_{sa}$ | 0.7% |
| f(sep.) | 6.0% | 17.8% | 11.8% | | $N_{sa}$ | 0.12 |

**Supporting Information**

In this section the derivation of the collision outcome model [8] and the post-separation characteristics model [9] for droplet-droplet interactions is provided.

## S1.1 Reflexive separation for droplet-droplet collisions

Reflexive separation occurs during collisions close to head-on when the droplets coalesce and initially form a torus-like structure which then, if We is high enough, breaks up to form daughter droplets and sometimes satellite droplets, so that more than two droplets are created following the collision. Upon collision the droplets coalesce, and fractions of each droplet are stretched and experience counteractive and excess surface-induced flows with associated kinetic energies $KE_{ST}$, $KE_{CO}$ and $KE_{ES}$ combining to give the total reflexive energy $KE_R$. The equations describing these as derived by Ashgriz and Poo [8] are given as:

$$KE_{ES} = \sigma\pi d_2^2\left[(1+\Delta^2) - (1+\Delta^3)^{2/3}\right] \tag{S1}$$

$$KE_{CO} = \frac{\rho}{2}(V_{1i}\vec{u}_1^2 + V_{2i}\vec{u}_2^2) \tag{S2}$$

$$KE_{ST} = \frac{\rho}{2}[(V_1 - V_{1i})\vec{u}_1^2 + (V_2 - V_{2i})\vec{u}_2^2] \tag{S3}$$

$$KE_R = KE_{ES} + KE_{CO} - KE_{ST} \tag{S4}$$

resulting in an expression for the reflexive energy $KE_R$ (eq. S9) which is written purely in terms of the three dimensionless parameters already defined in eqs. (1-3) in the main text using the substitutions given in eqs. (S5-8):

$$V_{1i} = \frac{\pi}{6}d_2^3(\Delta - \xi)^2\sqrt{(\Delta^2 - \xi^2)}$$
$$V_{2i} = \frac{\pi}{6}d_2^3(1 - \xi)^2\sqrt{(1 - \xi^2)} \tag{S5}$$

$$\vec{u}_1 = \frac{\vec{u}_{rel}}{1+\Delta^3}, \qquad \vec{u}_2 = \frac{\Delta^3\vec{u}_{rel}}{1+\Delta^3} \tag{S6}$$

$$\eta_1 = 2(1-\xi)^2(1-\xi^2)^{1/2} - 1$$
$$\eta_2 = 2(\Delta-\xi)^2(\Delta^2-\xi^2)^{1/2} - \Delta^3 \tag{S7}$$

$$\xi = \frac{b}{2}(1+\Delta) \tag{S8}$$

$$KE_R = \sigma\pi d_2^2\left[(1+\Delta^2) - (1+\Delta^3)^{2/3} + \frac{We}{12\Delta(1+\Delta^3)^2}(\Delta^6\eta_1 + \eta_2)\right] \tag{S9}$$

So that the boundary condition for reflexive separation is set so that when the effective reflexive energy $KE_R$ exceeds 75% of the surface energy of the coalesced droplet, reflexive separation occurs:

$$KE_R \geq 0.75\sigma\pi(d_2^3 + d_1^3)^{2/3} \tag{S10}$$

And, combining eqs (S9) and (S10):

$$We = 3\Delta(1+\Delta^3)^2 \left[\frac{7(1+\Delta^3)^{2/3} - 4(1+\Delta^2)}{(\Delta^6\eta_1 + \eta_2)}\right]$$

(S11)

**S1.2 Stretching separation for droplet-droplet collisions [8]**

The second class of separations is the 'stretching' separation, occurring in the region of the regime plot occupied by separation in the O'Rourke formulation depicted in Figure 2 in the main text. Unlike the reflexive separation, where the two droplets coalesce and then separate, in the stretching separation only a portion of each droplet interacts and is exchanged during separation. There are therefore two components to the kinetic energy: that of the interacting portion and that of the non-interacting portion which continues on its initial trajectory. As with reflexive separation, the boundary between separation and coalescence is when the total kinetic energy exceeds the total surface energy of the interacting volume:

$$KE_S \geq SE_I$$

(S12)

The interacting fraction h of the two droplets derives from the difference between the sum of the drop radii and the impact parameter b, given as:

$$h = \frac{1}{2}(d_1 + d_2)(1-b)$$

(S13)

So that the interacting volumes for each droplet, $V_{1i}$ and $V_{2i}$, are given as:

$$V_{ni} = \phi_n V_n$$

(S14)

with $\phi_{1,2}$ defined as:

$$\phi_1 = \begin{cases} 1 - \frac{1}{4\Delta^3}(2\Delta - \tau)^2(\Delta + \tau) & \text{for } h > \frac{d_1}{2} \\ \frac{\tau^2}{4\Delta^3}(3\Delta - \tau) & \text{for } h < \frac{d_1}{2} \end{cases}$$

$$\phi_2 = \begin{cases} 1 - \frac{1}{4}(2 - \tau)^2(1 + \tau) & \text{for } h > \frac{d_2}{2} \\ \frac{\tau^2}{4}(3 - \tau) & \text{for } h < \frac{d_2}{2} \end{cases}$$

$$\text{where } \tau = (1-b)(1+\Delta)$$

(S15)

Thus, the kinetic energies of the non-interacting parts of the droplets are given by:

$$KE_{Sn} = \frac{\rho}{2}[(V_1 - V_{1i})\vec{u}_1^2 + (V_2 - V_{2i})\vec{u}_2^2]$$

(S16)

and the kinetic energies of the interacting parts are:

$$KE_{Si} = \frac{\rho}{2}[V_{1i}(\vec{u}_1 b)^2 + V_{2i}(\vec{u}_2 b)^2]$$

(S17)

so that the sum of the kinetic energies $KE_S = KE_{Sn} + KE_{SI}$ can be rewritten as:

$$KE_S = \frac{\rho}{2}\vec{u}_2^2 V_2^2 \left\{\frac{\Delta^3}{(1+\Delta^3)^2}[(1+\Delta^3) - (1-b^2)(\phi_1 + \Delta^3\phi_2)]\right\}$$

(S18)

The surface energy acting against the stretching kinetic energy is estimated as that of a cylinder with volume $V_{1i} + V_{2i}$ and height $h$ (see eqs. S13-15), neglecting the area of the two ends of the cylinder as they have no new interface with the droplets:

$$SE_I = \sigma[2\pi V_2 d_2 \tau(\Delta^3\phi_1 + \phi_2)]^{1/2}$$

(S19)

Combining eqs. (S12), (S18) and (S19) then gives the stretching separation boundary condition expressed in dimensionless terms:

$$We = \frac{4(1+\Delta^3)^2[3(1+\Delta)(1-b)(\Delta^3\phi_1 + \phi_2)]^{1/2}}{\Delta^2[(1+\Delta^3) - (1-b^2)(\phi_1 + \Delta^3\phi_2)]}$$

(S20)

### S1.3 Post-separation characteristics [9]

Ashgriz and Poo's model was extended to predict the sizes and velocities of separated droplets post-collision, as well as the production and characteristics of satellite droplets, by Ko and Ryou [9]. They introduced a new parameter, the separation volume coefficient $C_V$, to quantify the portion of the two colliding droplets that is separated during the collision and can form new satellite droplets. For both stretching and reflexive separations, equation (S21) gives $C_V$.

$$C_{V(R,S)} = \frac{KE - SE}{KE + SE}$$

(S21)

with the terms for KE and SE for reflexive and stretching separations given by eqs. (S9), (S10), (S18) and (S19). Ko and Ryou considered the state of the droplets pre- and post-collision from the perspective of conserved mass, momentum and energy, with the result that the diameters and velocities of the original droplets post-collision $d_{na}$ and $\vec{u}_{na}$ (where $n = 1,2$), and the number, diameter and velocity of any newly generated satellite droplets $N_{sa}$, $d_{sa}$ and $\vec{u}_{sa}$ are found to be:

| Stretching separation | Reflexive separation | |
|---|---|---|
| $d_{2a} = (1 - C_{VS}\phi_2)^{1/3} d_2$ | $d_{2a} = (1 - C_{VR})^{1/3} d_2$ | (S22a, b) |

$$\vec{u}_{na} = \vec{u}_n \qquad \boldsymbol{\vec{u}_{na}} = \frac{\sqrt{2}}{2}\boldsymbol{\vec{u}_n} \tag{S23a, b}$$

$$N_{sa} = \left[\frac{SE_S}{\sigma\pi C_{VS}^{2/3}(\Delta^3\phi_1 + \phi_2)^{2/3} d_2^2}\right]^3 \qquad N_{sa} = \left[\frac{SE_S}{\sigma\pi C_{VR}^{2/3}(d_1^3 + d_2^3)^{2/3}}\right]^3 \tag{S24a, b}$$

$$d_{sa} = \left[\frac{C_{VS}(\Delta^3\phi_1 + \phi_2)}{N_{sa}}\right]^{1/3} d_2 \qquad d_{sa} = \left[\frac{C_{VR}(d_1^3 + d_2^3)}{N_{sa}}\right]^{1/3} \tag{S25a, b}$$

$$\vec{u}_{sa} = \frac{\Delta^3(\phi_1 - \phi_2)}{(1 + \Delta^3)(\Delta^3\phi_1 + \phi_2)}\vec{u}_{rel} \qquad \vec{u}_{sa} = \frac{d_1^3\vec{u}_1 + d_2^3\vec{u}_2}{d_1^3 + d_2^3} \tag{S26a, b}$$

where $SE_S$ is the excess surface energy distributed between the satellite droplets and is calculated from the conservation of energy pre- and post-collision. Ko and Ryou compared the experimental data presented by Ashgriz and Poo on the observed number of satellite droplets formed ($N_{sa}$) and found good agreement with the predictions of their extended model.